\documentstyle[preprint,aps,prd]{revtex}
\input{epsf}
\tightenlines

\newcommand{\eq}{\begin{eqnarray}}
\newcommand{\en}{\end{eqnarray}}
\renewcommand{\theequation}{\arabic{section}.\arabic{equation}}
\newcommand{\co}{\; \; ,}

\newcommand{\scs}{\co \;}
\newcommand{\sem}{ \; \; ; \;}
\newcommand{\per}{ \; .}

\begin{document}
\preprint{BUTP-99/16, Bicocca-FT-99-25}

\draft
\title{Effective Lagrangians in bound state calculations}
\author{V.\ Antonelli$^{a,b}$,
        A.\ Gall$^a$\footnote{Present address: SWITCH, Limmatquai 138,
                8001 Z\"{u}rich},
        J.\ Gasser$^a$, and
        A.\ Rusetsky$^{a,c}$}

\address{$^a$ Institute for Theoretical Physics, University of Bern,
Sidlerstrasse 5, CH-3012, Bern, Switzerland}

\address{$^b$ Dipartimento di Fisica, Universit\'{a} di Milano e
Milano-Bicocca e INFN, Sezione di Milano,\\
Via Celoria 16 - 22100 Milano, Italy}

\address{$^c$ HEPI, Tbilisi State University, University St. 9,
380086 Tbilisi, Georgia}

\date{13 March 2000}

\maketitle

\begin{abstract}
In order to investigate  specific aspects  of bound state
calculations in a non-relativistic framework, we consider the
energy-levels of a massive scalar particle, which moves in an
external field and interacts in addition with a massless
 scalar particle. The discussion includes the following topics:
dimensionally regularized bound-state calculations, ultraviolet
finiteness of bound-state observables and their  independence of
the off-mass-shell behavior of Green functions,  non-renormalizable
 interactions, structure of the non-relativistic two-point function,
power counting and matching.
\end{abstract}


\tableofcontents

\section{Introduction}

In several recent publications~\cite{pipi-NR,Bern1,Bern2}, the decay of
$\pi^+\pi^-$
atoms has been studied in the framework of a non-relativistic effective
Lagrangian - a method originally proposed by Caswell and Lepage~\cite{Lepage}
to investigate bound states in general. This method has proven to be
far more efficient for the treatment of loosely bound systems - such
as the $\pi^+\pi^-$ atom - than conventional approaches based on
relativistic bound-state equations~\cite{pipi-rel}. In~\cite{Bern1}, we have
discussed the chiral expansion of the decay width in this framework. The
corresponding numerical analysis is carried out in~\cite{Bern2}.

While performing that investigation, we have been confronted with
 the following specific features of this approach:
\begin{itemize}
\item[i)]
The non-relativistic effective Lagrangian defines a non-renormalizable
Quantum Field Theory, where Green functions are expanded in
powers of the coupling constant and of external momenta. While the expansion
in the coupling constant can be  done in the standard manner, the expansion
in momenta is highly non-trivial: In order to make sense of the perturbative
expansion,  one needs a procedure where  the contributions of
higher dimensional operators in the loop expansion are suppressed.
This problem goes under the name ``power counting''.
\item[ii)]
The renormalization program in this non-relativistic theory
must be carried out in a consistent manner. A reduction formula should
be available, that allows one to evaluate scattering matrix elements
from the residues of off-shell Green functions, analogous to the
procedure used in the relativistic formulation. For this, the
structure of the two-point function in the non-relativistic theory
must be known.
\item[iii)] One needs a systematic framework to calculate energy-levels and
decay widths in the non-relativistic theory. In view of  applications in
hadronic atom calculations, the framework should allow one to also
evaluate the lifetime of unstable states in an unambiguous manner.
\item[iv)]
The effective Lagrangian contains terms with arbitrary high powers of
space derivatives. As a result of this, the matrix elements of the
perturbation Hamiltonian between the unperturbed wave functions start to
diverge at a sufficiently high order of the perturbative expansion. One
should  demonstrate that these divergences cancel at any given
order in the expansion.
\item[v)]
The scattering sector of $\pi\pi$ interactions is most effectively
described by Chiral Perturbation Theory (ChPT). Here, the ultraviolet (UV)
divergences are removed by higher
order counterterms in the chiral expansion. One  needs an analogous
statement concerning the ultraviolet divergences in bound-state observables.
 \end{itemize}

Some of these issues - like the problem of power counting in the context of
QED bound states~\cite{QED,Soto,Czarnecki}, NRQCD and
HQET~\cite{NRQCD-HQET,Manohar,Beneke,Griesshammer}, effective field theories in
the one~\cite{GSS,HBChPT,Becher} and two-nucleon sectors~\cite{NN}
- have been treated previously in the literature, while some others have not
yet been, as far as we are aware, considered in detail. In particular,
 a detailed discussion of the renormalization program in the
non-relativistic sector - including an investigation of the structure
of the two-point function -  is not yet available to the
best of our knowledge. For this reason, we present
 in this article two examples of  Quantum Field Theories, where
 these issues  can be studied in an unambiguous manner.

The article is organized as follows. In section~\ref{sec:rel}, we present the
relativistic formulation of the models: First, we consider a massive scalar
particle moving in a given external field. Second, we add a Yukawa
interaction of the heavy particle with a massless scalar particle. This
interaction changes the energy-levels - we evaluate the shift at lowest
non-trivial order in the ground state. This section also contains an example
of a non-renormalizable interaction, that nevertheless results in
ultraviolet finite
shifts of the energy-levels. In section~\ref{sec:NRex}, we formulate the
non-relativistic
version of the external field problem. A general formula for evaluating
bound-state poles is derived with the use of Feshbach's~\cite{Feshbach}
projection technique. We determine the perturbative expansion of the
energy-levels, using dimensional regularization to tame the ultraviolet
divergences
that occur in these calculations. In section~\ref{sec:dynamical}, we
discuss the inclusion of a dynamical light field in the
non-relativistic framework, and illustrate the problem generated by
loop corrections.
The section~\ref{sec:twopoint} is devoted to a detailed study of the
two-point function in the non-relativistic theory, while the matching
procedure, needed to determine the coefficients in the
non-relativistic Lagrangian, is considered in
section~\ref{sec:matching}.
 In section~\ref{sec:groundstate}, we evaluate the energy-level shift in
the ground state and explicitly verify that the result agrees with the
relativistic calculation performed in section~\ref{sec:rel}.  Finally,
section~\ref{sec:summary}
contains a summary and concluding remarks. The Appendices are devoted to
technical details of the calculations presented in the text.

\setcounter{equation}{0}
\section{Relativistic formulation}
\label{sec:rel}
We investigate in this section the bound state spectrum of a heavy
particle,
which is bound in an external Coulomb field, and which interacts in
addition with a
massless scalar particle.  The results of the relativistic formulation
outlined here are compared in  later sections with a non-relativistic
framework.
\subsection{The external field}
We consider a heavy, complex scalar field $H$ of mass $M$, interacting with
an external, static source $\phi$,
\eq\label{eq:2-01}
  {\cal L}=\partial_\mu H^\dagger\partial^\mu H -M^2 H^\dagger H
  +\lambda H^\dagger H\phi\per
\en
To determine the energy spectrum, one may proceed in several ways.  In the
following, we consider the connected two-point function
\eq\label{eq:2-02}
  \bar G(x,y)=i\langle 0|TH(x)H^\dagger(y)|0\rangle_c\co
\en
because its Fourier transform contains the bound state poles.  It is useful to
note that $\bar G$ is the inverse of a differential operator,
\eq\label{eq:2-03}
  (D_M-\lambda \phi)\bar G(x,y) = \delta^4(x-y)\sem D_M=\Box_x+M^2\co
\en
as a result of which one has\footnote{
We use throughout the Landau symbols $O(x)$ [$o(x)$] for quantities that
vanish like $x$ [faster than $x$] when $x$ tends to zero. Furthermore,
it is understood that this holds modulo logarithmic terms, i.e. we write also
$O(x)$ for $x\ln x$.
}
\eq\label{eq:2-04}
  \bar G&=& D_M^{-1} +D_M^{-1}TD_M^{-1}\co\nonumber\\[2mm]
  T&=& (1-\lambda \phi D_M^{-1})^{-1}\lambda
  \phi=\lambda\phi+\lambda^2\phi D_M^{-1}\phi+O(\lambda^3)\per
\en
The scattering amplitude is obtained from the Fourier transform of $T$ by
putting the momenta on the mass-shell,
\eq\label{eq:2-05}
  \int dx^0 d^3{\bf x}d^3{\bf y} e^{ip^0(x^0-y^0)}e^{-i{\bf p}{\bf x}}
  e^{i{\bf k}{\bf y}}T(x,y)  = T({\bf p},{\bf k})\co
\en
with
\eq\label{eq:2-06}
  p^2=M^2\scs |{\bf p}|=|{\bf k}|\per
\en
Expanding in powers of $\lambda$, we find
\eq\label{eq:2-07}
  T({\bf p},{\bf k}) = \sum_{n=1}^\infty \lambda^n\int
  K_n({\bf p},{\bf k}; {\bf q}_1,\ldots,
{\bf q}_n)\prod_{i=1}^n\phi({\bf q}_i) d\nu({\bf q}_i)\co
\en
with
\eq\label{eq:2-08}
  \phi({\bf q})=\int d^3{\bf{x}} e^{-i{\bf q} {\bf x}} \phi({\bf x})\, ,
\quad\quad d\nu({\bf q})=d^3{\bf q}/(2\pi)^3\per
\en
The kernels $K_n$ are independent of the mass $M$,
\eq\label{eq:2-09}
  K_1({\bf p},{\bf k}; {\bf q}_1) &=&
  (2\pi)^3\delta^3({\bf q}_1-{\bf p}+{\bf k})\, ,\nonumber\\[2mm]
  K_2({\bf p},{\bf k};{\bf q}_1,{\bf q}_2)&=&(2\pi)^3
  \delta^3({\bf q}_1+{\bf q}_2-{\bf p}+{\bf k})
  \frac{1}{({\bf k}+{\bf q}_2)^2-{\bf k}^2}\per
\en
In the following, we often consider the case where the external field is of
the Coulomb-type,
\eq\label{eq:2-10}
  \phi=\frac{1}{|{\bf x}|}\, .
\en
Below, we shall not always distinguish between the case
of  short and long range (Coulomb) external sources. In the
latter case, the on-shell scattering
amplitude contains infrared divergences that are tamed by using
dimensional regularization.

The Fourier transform of the propagator
\eq\label{eq:2-11}
  \bar G(E;{\bf p},{\bf k})=\int dx^0 d^3{\bf x}d^3{\bf y}e^{iE(x^0-y^0)}
  e^{-i{\bf p}{\bf x}}e^{i{\bf k}{\bf y}}\bar G (x,y)
\en
is then closely related to Schwinger's Green function~\cite{Schwinger},
\eq\label{eq:2-12}
  \bar G(E;{\bf p,k}) = -\frac{(2\pi)^3}{2M}\,\,
  G_S\biggl(\frac{E^2-M^2}{2M};{\bf p,k}\biggr)_{Ze^2=\frac{\lambda}{2M}}
  \per
\en
The explicit expression is given by\footnote{In order to simplify the
  notation, we omit the negative imaginary part in $M^2$ that is understood
  here and in the following.}
\eq\label{eq:2-13}
  \bar G(E;{\bf p},{\bf k}) &=& -\frac{(2\pi)^3\delta^{(3)}({\bf p}-{\bf k})}
  {E^2-M^2-{\bf p}^2}+\frac{1}{E^2-M^2-{\bf p}^2}\,
  \frac{4\pi\lambda}{|{\bf p}-{\bf k}|^2}\,\frac{1}{E^2-M^2-{\bf k}^2}
  \nonumber\\[2mm]
  &+& \frac{1}{E^2-M^2-{\bf p}^2}\,\,
  4\pi\lambda\nu I(E;{\bf p},{\bf k})\,\,\frac{1}{E^2-M^2-{\bf k}^2}\, ,
\en
with
\eq\label{eq:2-14}
I(E;{\bf p},{\bf k})=
\int_0^1\frac{\rho^{-\nu} d\rho}
  {[({\bf p}-{\bf k})^2\rho+\nu^2/\lambda^2(1-\rho)^2(M^2+{\bf p}^2-E^2)
    (M^2+{\bf k}^2-E^2)]}\co
\en
where $\nu=\frac{1}{2}\,\lambda\,(M^2-E^2)^{-1/2}$.

In the representation (\ref{eq:2-13}), the two-point function is split into
three
parts. The two first terms correspond to free propagation and to one
interaction with the external field, respectively, whereas the third term
describes multiple interactions. The
function $\bar G(E;{\bf p},{\bf k})$ develops poles at $\nu=n$ or,
equivalently, at
\eq\label{eq:2-15}
    E_n^2=M^2-\frac{\lambda^2}{4n^2}\per
\en
The residues factorize in a product of two wave functions which are solutions
of the Schr\"{o}dinger equation with a Coulomb potential,
\eq\label{eq:2-16}
  \left[E_n^2-M^2-{\bf p}^2\right]\psi_n({\bf p})=
  -4\pi\lambda\int\frac{d\nu({\bf q})}{|{\bf p}-{\bf q}|^2}\psi_n({\bf
    q})\per
\en
For later use, we display the normalized ground-state wave function in
momentum space,
\eq\label{eq:2-17}
  \psi_1({\bf p})=
  \frac{(64\pi\gamma^5)^{1/2}}{({\bf p}^2+\gamma^2)^2}\co\gamma=\lambda/2\sem
\quad\quad  \int d\nu ({\bf p})\, |\psi_1({\bf p})|^2=1\per
\en

\subsection{Radiative corrections}

We now supplement the Lagrangian (\ref{eq:2-01}) with a Yukawa-interaction
of the heavy particle with a massless scalar field.
 This additional  interaction
affects the energy-levels (\ref{eq:2-15}) - we shall investigate
the induced change in $E_1^2$ at lowest non-trivial order in the Yukawa
coupling.

\subsubsection{The Lagrangian}
The Lagrangian  is
\eq\label{eq41}
{\cal L}&=&
\partial^\mu H^\dagger\partial_\mu H-M^2H^\dagger H+
\frac{1}{2}(\partial^\mu\ell\partial_\mu\ell-m^2\ell^2)
+H^\dagger H(\lambda\phi+e\ell)+{\cal L}_{c.t.}\co\nonumber\\
{\cal L}_{c.t.}&=&
-\delta M^2H^\dagger H-\frac{1}{2}\delta m^2\ell^2
+(c_0+c_1\lambda\phi+c_2\lambda^2\phi^2)\ell\per
\en
The mass $m$ is used as an infrared regulator in intermediate steps.
Whenever possible, we always consider the limiting case
$m\rightarrow 0$.
The counterterms - collected in the quantity ${\cal L}_{c.t.}$ -
 cancel the ultraviolet divergences in the diagrams displayed in
Fig.~\ref{UV}. As is discussed below, the
counterterm proportional to $c_2$ is needed
in order to  have a smooth behavior of the scattering amplitude in
the limit $m\rightarrow 0$.

In dimensional regularization, we find
\eq\label{eq49}
&&\delta M^2=-e^2\bar L_M +\frac{e^2}{8\pi^2}+O\biggl(\frac{m}{M}\biggr)\co
\quad
\delta m^2=-e^2\bar L_M+O\biggl(\frac{m^2}{M^2}\biggr)\co
\nonumber\\[2mm]
&&c_0=-eM^2\bar L_M+\frac{eM^2}{16\pi^2}\co
\quad
c_1=e\bar L_M\co
\quad
c_2=-\frac{e}{32\pi^2M^2}\co
\en
where $D$ stands for the dimension of the space-time, and
\eq\label{eq48}
\bar L_M&=&\frac{1}{8\pi^2}
\biggl(\bar L(\mu)+\frac{1}{2}\ln\frac{M^2}{\mu^2}\biggr)\co
\nonumber\\[2mm]
\bar L(\mu)&=&
\mu^{D-4}\,\biggl(\frac{1}{D-4}-\frac{1}{2}(\Gamma'(1)+\ln 4\pi)\biggr)\, ,
\en
where $\mu$ denotes the renormalization scale.
The finite parts of the mass counterterms are chosen such that $M$ coincides
with the physical mass of the heavy particle at order $e^2$, and that
the light field stays massless at this order.
As is discussed below, the  finite parts
of the couplings $c_i$ as chosen in (\ref{eq49}) guarantee a smooth
behavior of the scattering amplitude in the limit $m\rightarrow 0$.
With (\ref{eq49}), the Green functions of the heavy and light
fields are ultraviolet finite at order $e^2$ and at any order in $\lambda$.

\subsubsection{Perturbation theory for bound states}
The Yukawa interaction (\ref{eq41})
generates a shift in the energy-levels (\ref{eq:2-15}).
 In order to derive the
 formula that allows us to calculate this shift, we start from the
 connected two-point function of the heavy scalar field in
 (\ref{eq41}),
\eq
 G(x,y)=i\langle 0|TH(x)H^\dagger(y)|0\rangle_c\per
\en
It obeys the familiar Dyson-Schwinger equation
\eq\label{eq:3-02}
  G(x,y)=\bar G(x,y)+\int dudv\, \bar G(x,u)\Sigma(u,v)G(v,y)\co
\en
where $\bar G$ stands for the Green function (\ref{eq:2-02}) in the
 absence of the Yukawa interaction. The relation (\ref{eq:3-02}) defines
the self-energy  $\Sigma$ of the heavy scalar particle.
  The expansion of $\Sigma$ in powers of the coupling
constant $e$ takes the form
\eq\label{eq:3-03}
  \Sigma(u,v)=\Sigma_{2}(u,v)+\Sigma_{4}(u,v)
  +\cdots\, ,
\en
where $\Sigma_n=O(e^n)$. Explicitly, we find for the lowest order expression
\eq\label{eq410}
\Sigma_{2}(u,v)&=&e\delta(u-v)\int dw\, D_m^{-1}(u-w)\,
[-ie\bar G(w,w)+c_0+c_1\lambda\phi(w)+c_2\lambda^2\phi^2(w)]\nonumber\\[2mm]
&-&\delta M^2\delta(u-v)
-ie^2\, D_m^{-1}(u-v)\bar G(u,v)\co
\en
where $D_m^{-1}(x)$ denotes the propagator of the free
 two-point function of the light field,\\
$(\Box+m^2)D_m^{-1}(x)=\delta^4(x)$.

It is convenient to define the Fourier transforms
\eq
\Sigma(E;{\bf x},{\bf y})=\int dx^0 e^{iE(x^0-y^0)}\,
\Sigma(x,y)\co
\en
and
\eq
\Sigma(E;{\bf p},{\bf k})=\int d^3{\bf x}d^3{\bf y}
\, e^{-i{\bf px}+i{\bf ky}}\Sigma(E;{\bf x,y})\per
\en

The diagrams contributing to $\Sigma_2$  are depicted in
Fig.~\ref{I-II}. We have not displayed the contributions
generated by the counterterms $c_i$ and $\delta M^2$.
There are two types of diagrams: the tadpole-like diagram I, where
the loop is generated by the propagator $\bar{G}$,
and the self-energy diagram II, generated by  a closed loop with
the light scalar and $\bar{G}$.
The tadpole-like diagram leads to a self-energy of the form
\eq
\Sigma_2^I(E;{\bf p},{\bf k})=e^2\frac{R({\bf q})}{m^2+{\bf q}^2}\sem
{\bf q}= {\bf p}-{\bf k}\co
\en
where the residue behaves at low momentum transfer as
\eq
R=R_0\delta^3({\bf q}) +\frac{R_1}{{\bf q}^2}
+\frac{R_2}{|{\bf q}|}+R_3\ln{\frac{|{\bf q}|}{M}}+O(1)\scs {\bf
q}\rightarrow 0\co
\en
and where $R_m=O(\lambda^m)$. The finite parts in the counterterms are
chosen such that $R_{0,1,2}=0$. As a result of this, the regulator
mass $m$ can be sent to zero, and the self-energy is not more singular
(except from logarithms) than the external Coulomb-field in the
forward region ${\bf p}={\bf k}$. In particular, one has
\eq
\Sigma_2^I(E;{\bf p},{\bf k})
=-\frac{\lambda e^2}{24\pi M^2}\frac{1}{{\bf q}^2}
\left\{ 1+O({\bf q}^2)\right\}+O(e^2\lambda^2)\per
\en
The coupling to the external Coulomb-field is modified accordingly
at this order,
\eq\label{eq416}
\lambda\rightarrow\lambda\left[1-\frac{e^2}{96\pi^2M^2}\right]\per
\en
 Note  that in QED,  the coefficients $R_{0,1,2}$ are identically zero
 by charge conjugation and gauge invariance.

Coming back to the bound-state energy, we observe that the Fourier transform
of $G(x,y)$ has a pole at the
energy $E_B$ corresponding to the bound state $B$,
\eq\label{eq:3-04}
  G(E;{\bf x},{\bf y})=\int dx^0 e^{iE(x^0-y^0)}\,
  G(x,y)
\rightarrow  -\,\frac{\langle 0|H(0,{\bf x})|B\rangle
    \langle B|H^\dagger(0,{\bf y})|0\rangle}{2E_B(E-E_B)}+\cdots\per
\en
Using Eqs.~(\ref{eq:2-03}), (\ref{eq:3-02}) and
(\ref{eq:3-04}), one obtains for the bound-state wave function,
defined as $\psi({\bf x})=\langle 0|H(0,{\bf x})|B\rangle$,
the  equation
\eq\label{eq:3-05}
  \bigl(\, E_B^2-M^2+\triangle_x+\lambda\phi({\bf x})\,\bigr)\,\psi({\bf x})=
  -\int d^3{\bf y}\,\Sigma(E_B;{\bf x},{\bf y})\,\psi({\bf y})\, .
\en

 In order to calculate the corrections to the energy-levels,
one can apply ordinary perturbation theory to the Schr\"{o}dinger-like
equation~(\ref{eq:3-05}).
The  result is
\eq\label{eq:3-06}
  \Delta E_n^{~2}=
\bar E_n^2-E_n^2=-\int d\nu({\bf p})\, d\nu({\bf k})\,\,
\psi_n^*({\bf  p})\, \Sigma_2(E_n;{\bf p},{\bf k})\,\psi_n({\bf k})+O(e^4)
\, .
\en
Here $\psi_n$ denotes the normalized unperturbed Coulomb eigenfunction
(\ref{eq:2-16}), whereas $\bar E_n^2$ and $E_n^2$ stand for the
true and unperturbed eigenvalues, respectively~\cite{Bogoliubov}.

\subsubsection{The ground state}

We are now ready to evaluate the shift in the energy-levels, generated
by the Yukawa interaction $eH^\dagger H l$ in (\ref{eq41}). We
restrict ourselves to the calculation of the shift in the
ground-state. By dimensional reasons, one has
\eq\label{eq421}
\Delta E_{1}^2=e^2F(z)+O(e^4)\, ,\quad\quad z=\lambda/M \per
\en
Below, we  determine the leading term in the expansion of $F$ in powers of
$z$.
 The calculation can be performed at
zero mass of the light particle, because the bound particle is always slightly
off-shell - there is no additional infrared regulator needed.
The shift is naturally split into the two parts originating from the
diagrams of type I and II displayed in Fig.~\ref{I-II}.
We split the contributions from the diagrams of  type II
 furthermore into a zero-Coulomb,
 a one-Coulomb and a multi-Coulomb part according to
Eq.~(\ref{eq:2-13}), and write
\eq
F=F_I+F_{II,0}+F_{II,1}+F_{II,m}\per
\en
The contribution from the diagrams of the type I is remarkably
simple to evaluate.
Indeed, the previous discussion has shown that -
  at  leading order in $\lambda$ - the
inclusion of these diagrams merely amounts to an $O(e^2)$ correction
 in  the strength of the Coulomb potential,
  see Eq.~(\ref{eq416}). The contributions of order $\lambda^2$
  and higher in $\Sigma_2^I$ may be discarded at
leading order. One therefore has
\eq\label{eq422}
F_I=\frac{z^2}{192\pi^2}+o(z^2)\per
\en
The expression for the zero-Coulomb part in the diagram II
 starts again at order $z^2$, modulated with a
 logarithmic term,
\eq\label{eq425}
F_{II,0}=-\frac{z^2}{32\pi^2}\,\biggl(\ln z^2-\frac{1}{2}\biggr)+o(z^2)\, .
\en
The one-Coulomb  contribution  is given by
\eq\label{eq426}
F_{II,1}=-\int d\nu({\bf p})
 d\nu({\bf k})\,\,\psi_1^*({\bf p})\,\,
\frac{4\pi\lambda}{|{\bf p}-{\bf k}|^2}\,\,
\Gamma(E_1;{\bf p},{\bf k})\,\,\psi_1({\bf k})\co
\en
where $\Gamma$ stands for the one-loop correction to the coupling of the
heavy particle to the external field,
\eq\label{eq427}
&&\Gamma(E_1;{\bf p},{\bf k})
=-\int\frac{d^4q}{(2\pi)^4i}\,\,\frac{1}{q^2}\,\,
\frac{1}{(M^2-(E_1-q^0)^2+({\bf p}-{\bf q})^2)
(M^2-(E_1-q^0)^2+({\bf k}-{\bf q})^2)}\per
\nonumber\\
&&
\en
The calculation of the one-Coulomb contribution to the energy shift is
relegated to Appendix~\ref{app:1-C}. We simply quote the final result,
\eq\label{eq430}
F_{II,1}=\frac{z^2}{32\pi^2}\, (\ln z^2-6+8\ln 2)+o(z^2)\per
\en
The logarithmic terms cancel in the sum of  the zero- and one-Coulomb
 contributions.

The calculation of the multi-Coulomb contribution is given in
Appendix~\ref{app:multi-C}.
The result can be written as
\eq\label{m-C-J}
F_{II,m}=-\frac{z^2}{32\pi^2}\,\, J+o(z^2)\co
\en
with
\eq\label{J}
J&=&256\pi^2\,
\int_0^\infty\frac{xdx}{(1+x)^{1/2}}\int d\nu({\bf u})
 d\nu({\bf v})
\int_0^1\frac{d\rho\rho^{-(1+x)^{-1/2}}}
{(1+{\bf u}^2)^2(1+{\bf v}^2)^2(1+x+{\bf u}^2)(1+x+{\bf v}^2)}\times
\nonumber\\[2mm]
&\times&\frac{1}
{({\bf u}-{\bf v})^2\rho+(1-\rho)^2(1+x+{\bf u}^2)(1+x+{\bf v}^2)/(4(1+x))}
\, .
\en
Collecting everything, we find
\eq\label{eq437}
F(z)=-\frac{z^2}{32\pi^2}
\biggl(\,\frac{16}{3}-8\ln 2+J\biggr)+o(z^2)\per
\en

\subsection{Non-renormalizable interactions}

According to Eq.~(\ref{eq:3-06}),
 the corrections to the Coulomb energy-spectrum are
obtained at leading order by folding the self-energy with the
 unperturbed Coulomb wave functions in momentum space. In theories
 with non-renormalizable interactions - e.g. in ChPT
- the amplitudes grow faster in the external momenta at
 every order in the loop expansion, and the just mentioned integration
 starts to diverge at some order. This divergence is not, at first
 glance, cured by the counterterms in the Lagrangian,
 because it arises from an additional integration of the matrix
element that was already made finite by these counterterms.
In the context of the $\pi^+\pi^-$-atom, it has been claimed
 in Ref.~\cite{Bunatian} that, for this reason, one needs to
 introduce an explicit ultraviolet cutoff in order to render the bound-state
observables finite. Here we wish to demonstrate in a simple model
 - bearing all
the important properties of the original problem - that this is not
 the case: the ultraviolet  divergences that occur at intermediate steps  in
the
 evaluation of the perturbed energy-levels are all  cancelled.

In order to illustrate the point, we consider the Lagrangian
\eq\label{eq:3-07}
  {\cal L}=\partial_\mu H^\dagger\partial^\mu H-M^2H^\dagger H
  +\lambda H^\dagger H\phi
+g\,\{(\Box+M^2)H^\dagger\}\,
  \lambda\phi\,\{(\Box+M^2)H\}+{\cal L}_{c.t.}\, ,
\en
where $\phi$ is a Coulomb field.
In the last term, we anticipate the necessity of adding  counterterms
 to cancel the ultraviolet divergences in the Green functions.

The self-energy part is again defined by (\ref{eq:3-02}), and
 the shift in the energy-levels can  be calculated from
 (\ref{eq:3-06}), with $e^2\rightarrow g$.  At  order
$g$,  we find for the self-energy part
\eq\label{eq:3-08}
  \Sigma(E;{\bf p},{\bf k})=\,(M^2-E^2+{\bf p}^2)\,\frac{4\pi\lambda g}
  {|{\bf p}-{\bf k}|^2}\,(M^2-E^2+{\bf k}^2)+\Sigma_{c.t.}(E;{\bf
p},{\bf k})\per
\en
The first term  is ultraviolet finite -
 however,
 it leads to an ultraviolet infinite contribution to the
energy shift at  order $g$. This fact mimics the situation which one
encounters in the calculation of the energy of hadronic atoms beyond the
so-called ``local''
approximation~\cite{Bunatian}. The remedy for the difficulty is provided by
the counterterm in Eq.~(\ref{eq:3-07}): as one may observe from
Eq.~(\ref{eq:3-02}), the two-point function [defined with the
 Lagrangian~(\ref{eq:3-07})]  is infinite at order $g$,
although the self-energy is not.
In dimensional regularization, the ultraviolet divergent piece
in the two-point function is given by
\eq\label{eq:3-09}
  G(E;{\bf p},{\bf k})\biggr|_{div}=
  -2\pi\lambda^3g L(\mu)\,
  \int d\nu({\bf q})d\nu({\bf l})
\bar G(E;{\bf p},{\bf q})\,\,  \bar G(E;{\bf l},{\bf k})+O(g^2)\, ,
\en
where $L(\mu)$ is given by
\eq\label{Lmu}
L(\mu)=(\mu^2)^{D-4}\biggl(\frac{1}{D-4}-\Gamma'(1)-\ln 4\pi\biggr)\per
\en
This divergence is cancelled by the counterterm
\eq\label{eq:3-10}
{\cal L}_{c.t.}=-\frac{1}{2}\,\lambda^3g\,c\,
H^\dagger(x)H(x)\triangle\phi({\bf x});\quad\quad
c=L(\mu)+c^r(\mu)\co
\en
where $c^r(\mu)$ denotes the arbitrary scale-dependent finite part of $c$.
 If one would carry out the perturbative expansion of
 the two-point
function to higher orders in  $g$, one would have to
  introduce additional counterterms, containing higher
derivatives - the situation
clearly resembles the one in ChPT.
Evaluating $\Sigma_{c.t.}$ generated by the counterterm (\ref{eq:3-10}),
 one can  verify that the corresponding contribution
 to the energy-level shift cancels the ultraviolet divergence generated by the
first
 term in (\ref{eq:3-08}),
 leading to a  finite and scale-independent result  at order $g$.

Let us note  that the ultraviolet divergences can be
removed from the bound-state observables in a perturbative manner:
indeed,
the divergent piece of the two-point function (\ref{eq:3-09})
was calculated at $O(g)$. The corresponding counterterm   removes
the ultraviolet divergence
 in the energy-levels at $O(g)$. This suggest that also
in non-renormalizable theories,  bound-state observables are
rendered finite at a given order, provided that the renormalization in
the scattering sector is carried out at the same order.

\setcounter{equation}{0}
\section{Non-relativistic formulation:\\The  external field}
\label{sec:NRex}
In the remaining part of this article, we
 wish to reformulate the relativistic problem considered in
section~\ref{sec:rel}
 in a non-relativistic
framework. The energy-levels can then be calculated using
Rayleigh-Schr\"{o}dinger perturbation theory.
At the same time, the formalism
 developed  allows one to calculate the width of decaying states
in a crystal clear manner, without using handwaving arguments.
 The procedure goes in two steps
\cite{Lepage}. First, one constructs an effective theory that allows
one to evaluate the scattering matrix elements in the low-energy region.
Second, one uses the Hamiltonian of this effective theory to evaluate the bound
state poles.

We restrict the considerations in this section to the non-relativistic
formulation of the external field problem (\ref{eq:2-01}).
 Radiative corrections generated by the Yukawa interaction
(\ref{eq41})  will be treated afterwards.

\subsection{The scattering sector}
For the construction of the non-relativistic effective Lagrangian that is
relevant for the
theory defined in Eq.~(\ref{eq:2-01}), one may proceed in several ways.
Below, we choose a method that is based on the following
observation. In the relativistic framework, the scattering matrix elements are
obtained from the residue of the poles in the Fourier transform of
the two-point function $\bar G$.
These poles are generated, in the present case, by the free propagators,
\eq\label{eq:2-18}
  D_M^{-1}(p)=\frac{1}{M^2-p^2}=
  \frac{1}{2\omega_{\bf p}}\left(\frac{1}{\omega_{\bf p}-p^0}
    +\frac{1}{\omega_{\bf p}+p^0}\right)\sem
  \omega_{\bf p}=(M^2+{\bf p}^2)^{1/2}\co
\en
which have poles at $p^0=\pm\omega_{\bf p} $. It is therefore sufficient to
construct a non-relativistic theory that has the same structure in the
vicinity of $p^0=\omega_{\bf p}$. The pole terms in (\ref{eq:2-18}) can be
generated with the inverse of the differential operators
\eq\label{eq:2-19}
  D_\pm &=&\pm i\partial_t -\sqrt{M^2-\triangle}\co \quad\quad
  D_\pm(x)\triangle_\pm(x)=\delta^{(4)}(x)\co
\nonumber\\[2mm]
  \triangle_\pm(x)&=&\int \frac{d^4p}{(2\pi)^4} \frac{e^{-ipx}}
  {\pm p^0-\sqrt{M^2+{\bf p}^2}}\per
\en
The Lagrangian will furthermore be quadratic in the non-relativistic heavy
field $h$. Writing ${\cal L}_{NR}=h^\dagger {\cal D}h$, the
differential operator
must ${\cal D}$ contain, in addition to $D_+$, further terms that generate the
transition amplitude $T$ in (\ref{eq:2-04}). Writing
\eq\label{eq:2-20}
  {\cal D}=D_+ + X\co\quad\quad
  {\cal D}^{-1}=D_+^{-1}-D_+^{-1}T_{NR}D_+^{-1}
\en
and requiring that $T_{NR}$ is closely related to the relativistic
scattering amplitude $T$ in (\ref{eq:2-04}), one finds \cite{Gall}
that one may take
\eq\label{eq:2-21}
  X = \lambda d \phi d(1+\lambda D_-^{-1}d\phi d)^{-1}
  \sem\quad\quad d\doteq (2\sqrt{M^2-\triangle})^{-1/2}\co
\en
with
\eq\label{eq:2-22}
  T_{NR}=dTd\per
\en
This formulation involves a non-local Lagrangian that generates the same
truncated off-shell two-point function as the relativistic theory.  In order
to arrive at a local Lagrangian, one may expand the non-local operators
$D_\pm^{-1}$ and $d$ in inverse powers of the heavy mass $M$. In this manner,
one arrives at a local Lagrangian, that contains, however, an infinite string
of terms.  Furthermore, it contains time derivative of arbitrary high orders,
acting on the heavy field. These time derivatives may be eliminated in a
standard manner by use of the equation of motion (EOM). We find that
this can be shown in a  transparent manner by use of
 external sources coupled to the
heavy field. We relegate these manipulations to Appendix~\ref{app:EOM}
and simply quote
the result,
\eq\label{eq:2-23}
  {\cal L}_{NR}&=&
  h^\dagger D_0 h +
  \sum_{n=0}^\infty\frac{1}{(2M)^{2n+1}}{\cal L}_{2n+1}\, , \nonumber\\[2mm]
  D_0&=& i\partial_t-M+\triangle/2M \co\nonumber \\[2mm]
  {\cal L}_1 &=&h^\dagger \lambda \phi h\co\quad
  {\cal L}_3 = h^\dagger \left[\triangle^2+\lambda(\phi\triangle +
\triangle\phi) +
    \lambda^2\phi^2 \right]h \co\nonumber\\[2mm]
  {\cal L}_5 &=& \frac{1}{2}h^\dagger \left[4\triangle^3+ \lambda
    \left(2\triangle\phi\triangle + {5}\phi\triangle^2 +
{5}\triangle^2\phi\right)
  \right]h\nonumber\\[2mm]
  &+& h^\dagger\left[{\lambda}^2 \left(2\phi^2\triangle \!+\!
      \triangle\phi^2 + 3\phi\triangle\phi\right) + 2{\lambda}^3\phi^3\right]
h\per
\en
The scattering matrix elements are obtained by evaluating the two-point
function
\eq\label{eq:2-25}
  G_{NR}(x,y)=i\langle 0|Th(x)h^\dagger(y)|0\rangle
\en
in the presence of the interactions ${\cal L}_{i}$.
The perturbation theory is
performed as follows. First one sums up all mass insertions in the external
lines. The Fourier transform of the two-point function then develops
poles at $p^0=\omega_{\bf p}$ and at $p^0=\omega_{{\bf k}}$,
\eq\label{eq:2-26}
G_{NR}(p^0;{\bf p},{\bf k})&=&  \int dx^0d^3{\bf x}d^3{\bf y}e^{ip^0(x^0-y^0)}
  e^{-i{\bf p}{\bf x}}e^{i{\bf k}{\bf y}}\, G_{NR}(x,y)
\nonumber\\[2mm]
&=&\frac{(2\pi)^3\delta^3({\bf p}-{\bf k})}{\omega_p-p^0}
+  \frac{1}{\omega_{\bf p}-p^0}\,
  \frac{1}{\sqrt{2\omega_{{\bf p}}}}\,\, T_{NR}({\bf p},{\bf k};p^0)\,\,
  \frac{1}{\sqrt{2\omega_{{\bf k}}}}\,\frac{1}{\omega_{{\bf k}}-p^0}\, .
\en
With the normalization chosen, $T_{NR}$ coincides with the low-energy
expansion of the relativistic amplitude $T$ in (\ref{eq:2-05}) order by
order. The presence of the factors $\sqrt{2\omega_{\bf p}}$,
$\sqrt{2\omega_{\bf k}}$ in the above formula reflects the
relation~(\ref{eq:2-22}).
The scattering amplitude is obtained by putting the momenta on the mass-shell
$p^0=\omega_{\bf p}$, $|{\bf p}|=|{\bf k}|$.  Expanding in powers of the
coupling constant, one has
\eq\label{eq:2-27}
  T_{NR}({\bf p},{\bf k}) = \sum_{n=1}^\infty \lambda^n\int
  k_n({\bf p},{\bf k}; {\bf q}_1,\ldots,{\bf
    q}_n;M)\prod_{i=1}^n\phi({\bf q}_i) d\nu({\bf q}_i)\per
\en
In the internal lines, one need not fully sum the mass insertions. Instead,
one sums, at a fixed order in $\lambda$, all terms at a given order in $1/M$,
\eq\label{eq:2-28}
  k_n({\bf p},{\bf k}; {\bf q}_1,\ldots,{\bf q}_n;M)
  =\sum_{\nu=0}^\infty\frac{1}{M^\nu}k_{n,\nu}
  ({\bf p},{\bf k}; {\bf q}_1,\ldots,{\bf q}_n)\per
\en
In Fig.~\ref{scat-ex} we display the corresponding graphs at
  order $\lambda^2$. The diagram Fig.~\ref{scat-ex}A stands for the Green
function
  evaluated in the relativistic theory, whereas  mass insertions
  and the contact term in the non-relativistic theory are displayed in
 graphs Fig.~\ref{scat-ex}B and Fig.~\ref{scat-ex}C.
The mass dependence cancels, order by
order, and the leading term coincides with the matrix element
in (\ref{eq:2-09}), evaluated in the
relativistic theory,
\eq\label{eq:2-29}
  k_{n,0}=K_n\sem k_{n,i}=0\co i \geq 1\per
\en
We illustrate the procedure with the matrix element at order $\lambda$. The
first two terms in the $1/M$ expansion give
\eq\label{eq:2-30}
  T_{NR}({\bf p},{\bf k}) = \lambda
  \phi({\bf p}-{\bf k})\frac{\omega_{\bf p} }{M}
  \left[1-\frac{{\bf p}^2}{2M^2}+O(M^{-3})\right]+O(\lambda^2)\per
\en
A different Lagrangian would be generated by use of the standard
Foldy-Wout\-huysen-Tani transformation, which gives
\eq\label{eq:2-31}
  {\cal L}_{FWT}=h^\dagger (i\partial_t-\left[M^2-\triangle-\lambda
    \phi\right]^{1/2}) h\per
\en
Again, the Lagrangian may be expanded in inverse powers of $M$. At order
$1/M^5$ and higher, it differs from the one considered above,
\eq\label{eq:2-32}
{\cal L}_{NR}-{\cal L}_{FWT} =
\frac{\lambda}{64M^5}h^\dagger\left\{
    \triangle^2 \phi -2\triangle \phi\triangle+\phi\triangle^2
    -2\lambda\triangle\phi^2+2\lambda\phi\triangle\phi\right\} h +\cdots\, .
\en
On the other hand, evaluating the scattering amplitude with (\ref{eq:2-31}),
one finds that it agrees with the previous one on the mass-shell, to all
orders in the coupling constant. The reason is the fact that the two
Lagrangians may be transformed into each other by a field redefinition. To
illustrate, the terms at order $1/M^5$ are removed if in ${\cal L}_{NR}$ we
substitute
\eq\label{eq:2-33}
  h&\rightarrow& (1+X)h\, ,
\nonumber\\[2mm]
  X&=& -\frac{1}{32M^4}\left[\lambda(\triangle \phi -\phi\triangle)
    -\lambda^2\phi^2\right]\per
\en
On the other hand, the two-point functions differ off the mass-shell.
Indeed, consider the term linear in $\lambda$. From (\ref{eq:2-32}) it follows
that the difference is proportional to\\
$\lambda\phi({\bf p}-{\bf k})({\bf p}^2 -{\bf k}^2)^2$ at order $M^{-5}$.

\subsection{Perturbation theory for bound states}

In Quantum Mechanics, bound states are often treated by splitting
the Hamilton operator
into a part with known  spectrum and known wave functions, and a
perturbation whose
effect can be systematically calculated in a power series in some small
parameter. This procedure is successful, because all truly
non-perturbative effects are collected in a compact manner in the lowest
order wave functions. In Quantum  Field Theory, however, the basic
quantities are the
Green functions, and there is no simple analogue to wave functions.
This makes the treatment of bound states inherently more complicated.  The
salient
feature of the effective non-relativistic theory is that it allows for a
consistent Hamiltonian formulation in the context of a field theory.

To see how this works, we define the Hamiltonian pertaining to the
Lagrangian~(\ref{eq:2-23}),
\eq
  &&{\cal H} = {\cal H}_0+{\cal H}_C+{\cal H}_I, \nonumber\\[2mm]
  {\cal H}_0= h^\dagger\left(M-\frac{\triangle}{2M} \right)h,
  \quad
  &&{\cal H}_C = -h^\dagger\frac{\lambda\phi}{2M}h , \quad
  {\cal H}_I = -\sum_{n=1}^\infty\frac{1}{(2M)^{2n+1}}{\cal L}_{2n+1}.
\en
The static external field preserves the time invariance of the theory.
We perform canonical quantization at $t=0$, introducing creation and
annihilation operators
\eq\label{aadagger}
{\bf a}^\dagger({\bf p})=\int d^3{\bf x}e^{i{\bf p}{\bf x}}
h^\dagger(0,{\bf x}),\quad
{\bf a}({\bf p})=\int d^3{\bf x}e^{-i{\bf p}{\bf x}}h(0,{\bf x})\, ,
\en
obeying the commutation relation $[{\bf a}({\bf p}),{\bf a}^\dagger({\bf q})] =
(2\pi)^3
\delta^{(3)}({\bf p}-{\bf q})$. The Hamilton operator is defined by ${\bf H} =
\int
d^3x {\cal H}(0,{\bf x})$ and similarly for ${\bf H}_0$, ${\bf H}_{\rm C}$ and
${\bf H}_{\rm I}$. A free
non-relativistic particle is described by the state $|{\bf p}\rangle =
{\bf a}^\dagger({\bf p})|0\rangle$, which obeys ${\bf H}_0|{\bf p}\rangle=
(M+{\bf p}^2/2M)|{\bf p}\rangle$.  The Coulomb wave functions $\psi_n$ are
defined as the
solutions of the non-relativistic Schr\"{o}dinger equation,
\eq\label{Schrodinger}
  \left( M+\frac{{\bf p}^2}{2M}\right)\psi_n({\bf p})
- \frac{2\pi\lambda}{M}
  \int d\nu ({\bf q})
  \frac{1}{|{\bf p}-{\bf q}|^2} \psi_n({\bf q}) = E_n\psi_n({\bf p}),
\en
with
\eq
  E_n = M-\frac{M}{2n^2}\left(\frac{\lambda}{2M}\right)^2\,
\en
for the discrete spectrum.
Note that the wave functions (\ref{eq:2-16}) satisfy the
equation~(\ref{Schrodinger}).

The eigenstates $({\bf H}_0 +{\bf H}_{\rm C})|n\rangle=E_n|n\rangle$ of the
unperturbed
Hamilton operator are given by
\eq\label{states}
  |n\rangle = \int d\nu ({\bf p})\psi_n({\bf p})|{\bf p}\rangle.
\en

Our task is to find the isolated poles  of the two-point
function~(\ref{eq:2-26}). Here we note that the spectrum
of the full theory is also encoded in the singularities of the resolvent
\eq
{\bf G}(E) = \frac{1}{E-{\bf H}}\per
\en
 If the Hamilton operator annihilates the vacuum,  ${\bf H}|0\rangle=0$,
the resolvent is closely related to the two-point function,
\eq
  \label{eq:resolvent-2pf}
  \langle{\bf p}|{\bf G}(E)|{\bf k}\rangle = -G_{NR}(E;{\bf p},{\bf k})\co
\en
see Appendix~\ref{app:resolvent}.
The advantage of using the resolvent instead of the Green
function is due  to the fact  that
the bound state poles can be isolated in a very convenient manner. As a lowest
order approximation, we ignore ${\bf H}_{\rm I}$. The resolvent
\eq
\bar{\bf G}(E) = \frac{1}{E-{\bf H}_0-{\bf H}_{\rm C}}
\en
of the unperturbed Hamilton operator is proportional to the Coulomb Green
function
given by Schwinger's solution,
\eq
   \langle{\bf p}|\bar{\bf G}(E)|{\bf k}\rangle=
(2\pi)^3\, G_S(E-M;{\bf p},{\bf k})\biggr|_{Ze^2=\frac{\lambda}{2M}}.
\en
If the perturbation is small, the poles of ${\bf G}(E)$ are expected to lie
in the vicinity of those of $\bar{\bf G}(E)$.
To work out the exact position of the poles, it is convenient
to first consider the operator ${\bf T}_{\rm I}$
defined through
\eq
{\bf T}_{\rm I}(E) = {\bf H}_{\rm I} +
{\bf H}_{\rm I}\bar{\bf G}(E){\bf T}_{\rm I}(E)\per
\en
It is related to the full resolvent by
${\bf G}(E)=\bar{\bf G}(E)+\bar{\bf G}(E){\bf T}_{\rm I}(E)\bar{\bf G}(E)$.
Applying a technique introduced a long time ago by Feshbach~\cite{Feshbach},
we use the projector $|n\rangle\langle n|$ to exhibit the pole near the
unperturbed value $E_n$,
\eq\label{eq:Tbar-pole}
{\bf T}_{\rm I}(E) =
               {\bf T}_n(E) +
\frac{{\bf T}_n(E)|n\rangle\langle n|{\bf T}_n(E)}
{E-E_n-\langle n|{\bf T}_n(E)|n\rangle},
\en
where ${\bf T}_n$ is defined by
\eq
\label{eq:definition-Tn}
{\bf T}_n(E) = {\bf H}_{\rm I} +{\bf H}_{\rm I}\bar{\bf G}_n(E){\bf T}_n(E)
\, ,\quad\quad
\bar{\bf G}_n(E)=\bar{\bf G}(E)({\bf 1}-|n\rangle\langle n|).
\en
The corresponding decomposition of the full resolvent is obtained by
expressing  ${\bf G}(E)$ through ${\bf T}_n(E)$,
\eq\label{Gpole}
&&{\bf G}(E)=\bar{\bf G}_n(E)+\bar{\bf G}_n(E){\bf T}_n(E)\bar{\bf G}_n(E)+
\frac{({\bf 1}+\bar{\bf G}_n(E){\bf T}_n(E))|n\rangle
\langle n|({\bf 1}+{\bf T}_n(E)\bar{\bf G}_n(E))}
{E-E_n-\langle n|{\bf T}_n(E)|n\rangle}.
\nonumber\\
&&
\en
The energy of the bound state is determined by the pole position $\bar E_n$
in the full resolvent.
Since both $\bar{\bf G}_n(E)$ and ${\bf T}_n(E)$ are regular in the vicinity
of $E_n$ that includes also $\bar E_n$, the pole position is given by
the zero of the denominator in the last term of
Eq.~(\ref{Gpole})~\cite{Feshbach},
\eq
  \label{eq:pole-position}
  \bar E_n -E_n-\langle n|{\bf T}_n(\bar E_n)|n\rangle = 0.
\en
So far, we did not make any approximations or pose restrictions on the
perturbation ${\bf H}_{\rm I}$. In fact the last formula holds even when we
add a ``radiation field'', see Section~\ref{sec:groundstate}. Moreover,
the same formula can be used for the analysis of the decays of quasistable
bound states~\cite{Bern1} - without introducing the concept of ``wave
function of the unstable state''.
The problem of determining the positions of the poles in the
two-point function $G_{NR}(E;{\bf p},{\bf k})$ is thus reduced to the
calculation of the matrix elements
 $\langle n|{\bf T}_n(E)|n\rangle$ in old-fashioned
 perturbation theory.

Having taken the non-perturbative part into account, it remains to solve
Eq.~(\ref{eq:pole-position}) in perturbation theory. In the case at hand,
$\lambda/2M$ is the only small parameter and we can express $\bar E_n$ as a
power series
\eq
  \bar E_n = E_n + \sum_{m=3}^\infty \lambda^m C_{m n}.
\en
The coefficients $ C_{m n}$ can be  obtained by using
 in (\ref{eq:pole-position}) the iteration
\eq\label{PT}
  \langle n|{\bf T}_n(\bar E_n)|n\rangle =
\langle n|{\bf H}_{\rm I}|n\rangle+
\langle n|{\bf H}_{\rm I}\bar{\bf G}_n(\bar E_n)
{\bf H}_{\rm I}|n\rangle + \cdots \co
\en
which can be calculated by inserting a complete set of
states - corresponding
to $ \bf{H_0}+ \bf{H_C}$ - and
keeping terms of the appropriate order in $\lambda$. Here, we recognize the
familiar Rayleigh-Schr\"{o}dinger perturbation theory.

\subsection{Energy-levels}

With a given Lagrangian, and a framework set, one can now calculate the
energy of the bound-state levels in the non-relativistic theory - the final
answer should reproduce of course the relativistic result. For example,
for the ground-state energy in the non-relativistic theory one should obtain
order by order in perturbation theory in the coupling constant $\lambda$
\eq\label{eq215}
E_1=\biggl( M^2-\frac{\lambda^2}{4}\biggr)^{1/2}=M-\frac{\lambda^2}{8M}
-\frac{\lambda^4}{128M^3}-\frac{\lambda^6}{1024M^5}+\cdots\, .
\en

Using Eq.~(\ref{PT}),
one encounters two conceptual difficulties.
First, as was demonstrated in the preceding discussion, the non-relativistic
Lagrangians which can be derived from the relativistic theory, are not unique,
although all of them yield the same $S$-matrix elements. In the case considered
here, they
differ by terms which vanish by the use of EOM
and field redefinitions.
Second, although all couplings in the non-relativistic
Lagrangian are finite, it contains terms with arbitrary high powers of
space derivatives. This means that the matrix elements of the perturbation
Hamilton operator between  the unperturbed wave functions start to diverge at a
sufficiently high order of the perturbative expansion. Consequently, one has
to demonstrate that: i) the terms which vanish by the use of EOM
 or which can be removed by field redefinitions, do not contribute to the
bound-state energy,
and ii) divergences in the bound-state energy cancel when all terms
at a given order in the coupling constant are summed up. We shall ensure that
this is what indeed occurs in explicit calculations.

In order to evaluate the energy-level shift, one has to specify the explicit
form of the Hamiltonian. We find it useful to start with the FWT
Hamiltonian, which has a remarkably compact form,
\eq\label{eq220}
{\cal H}_{FWT}=h^\dagger\,\sqrt{M^2-\triangle-\lambda\phi}\,\,\, h
=h^\dagger\biggl( M-\frac{\triangle+\lambda\phi}{2M}\biggr) h-
h^\dagger\biggl(\frac{(\triangle+\lambda\phi)^2}{8M^3}+\cdots\biggr) h\, .
\en
The first two terms in  ${\cal H}_{FWT}={\cal H}_0+{\cal H}_C+{\cal H}_I$
correspond to the unperturbed
non-relativistic Coulomb problem, and the rest is considered as a
perturbation. In order to regularize the ultraviolet divergences mentioned,
we perform the calculations in $d$ space dimensions [$d=D-1$], and perform the
limit
$d\rightarrow 3$ at the end. We define the Hamilton operator
${\bf H}_{\rm FWT}=\int d^d{\bf x}{\cal H}_{FWT}(0,{\bf x})$, and
similarly for ${\bf H}_{\rm 0}$, ${\bf H}_{\rm C}$ and ${\bf H}_{\rm I}$,
etc. Further, in $d$ dimensions, the momentum-space Coulomb wave
function $\psi^d({\bf p})$ is chosen to obey the $d$-dimensional
Schr\"{o}dinger equation with the same kernel as in Eq.~(\ref{Schrodinger}),
\eq\label{eq221}
\biggl( E_n^d-M-\frac{{\bf p}^2}{2M}\biggr)\,\psi_n^d({\bf p})=
-\frac{2\pi\lambda}{M}\,
\int d\nu_d({\bf q})\,\,\frac{1}{|{\bf p}-{\bf q}|^2}\,\,
\psi_n^d({\bf q})\, ,
\en
where $d\nu_d({\bf q})=d^d{\bf q}/(2\pi)^d$, and
$E_n^d$ denotes the energy of the $n$-th eigenstate in $d$ dimensions\footnote{
We have checked numerically that the ground state energy depends smoothly on
$d$ in the vicinity of physical space dimension $d=3$.
In particular, no bifurcations occur.}.
Using Eq.~(\ref{eq221}), it is seen that the
unperturbed Coulomb wave functions are eigenfunctions of
${\bf H}_{\rm I}$,
\eq\label{eq222}
\int d^d{\bf x}\,h^\dagger\,(\triangle+\lambda\phi)^k h\,
|n\rangle_d=(2M(M-E_n^d))^k\,|n\rangle_d\, ,
\en
where $|n\rangle_d$ is constructed with the use of Coulomb wave function
$\psi_n^d({\bf p})$, similarly to Eq.~(\ref{states}).
In this expression, one can safely put $d=3$. Using then the perturbation
series for the energy-levels, one  ensures that Eq.~(\ref{eq215})
holds in all orders in the coupling constant $\lambda$.

The simplicity of the above derivation is a consequence of the choice of
 the Hamiltonian in the FWT form~(\ref{eq220}) and should not be misleading.
 As we have already mentioned,  in general one encounters
ultraviolet divergences
 in  individual matrix elements in  sufficiently high orders of the
non-relativistic expansion.  To illustrate this fact, we again consider the
 original Lagrangian~(\ref{eq:2-23}).
The difference $\delta {\cal H}_I$ between the pertinent Hamiltonian
 (\ref{eq:2-23}) and the FWT Hamiltonian (\ref{eq220}) contains  terms at
$O(M^{-5})$ and higher,
\eq\label{eq225}
\delta {\cal H}_I=
-\frac{\lambda}{64M^5}\,h^\dagger
(\phi\triangle^2+\triangle^2\phi-2\triangle\phi\triangle) h
+\frac{\lambda^2}{32M^5}\, h^\dagger
(\triangle\phi^2-\phi\triangle\phi) h +O(M^{-7})\per
\en
It is expected that the level shift induced by the
 perturbation $\delta{\cal H}_I$  vanishes to all orders in
 $\lambda$. Consider matrix elements of the form
$_d\langle n|\int d^d{\bf x}\,h^\dagger (\triangle)^{k}
(\lambda\phi)^{l-k}h |m\rangle_d$,
with all possible permutations of the operators. Using
dimensional arguments, one finds that all such operators will
contribute at order
$O(\lambda^{2l})$ to the bound-state energy (up to possible
logarithms). Therefore, the energy shift induced by
 $\delta{\cal H}_I$ starts at  order $\lambda^6M^{-5}$ - this term  is
given by the diagonal matrix element of this operator
between the unperturbed  wave functions.
The following matrix element  is then e.g. needed,
\eq\label{eq223}\int d^d{\bf x}\, _d\langle n|\,
h^\dagger\, \triangle\phi\triangle\, h\,|m\rangle_d
=\int d\nu_d({\bf p})\, d\nu_d({\bf k})\,
\psi_n^d({\bf p})^*{\bf p}^2\,\frac{4\pi}{|{\bf p}-{\bf k}|^2}\,{\bf k}^2
\psi_m^d({\bf k})
\nonumber\\[2mm]
=-2\pi\lambda^2\psi^d_n({\bf 0})^*\psi_m^d({\bf 0})\,
\biggl( L(\mu)+\ln\frac{\lambda^2}{\mu^2}\biggr)+
O(1)\, ,\quad d\rightarrow 3\, ,
\en
where
\eq\label{eq224}
\psi_m^d({\bf 0})=\int d\nu_d({\bf p})\,\psi_m^d({\bf p}),
\en
and where $L(\mu)$ is given by Eq.~(\ref{Lmu}). Eq.~(\ref{eq223})
 illustrates the ultraviolet divergences mentioned.
Using the bound-state equation in $d$ dimensions~(\ref{eq221}) in
the calculation of all matrix elements from (\ref{eq225}),
one finds  that all divergent terms containing $L(\mu)$ cancel
 at order $\lambda^6 M^{-5}$. Moreover, it can be
 checked that the diagonal matrix element and,
therefore, the energy-level shift due to
$\delta{\cal H}_I$ vanishes at this order,
\eq\label{eq226}
_d\langle n|\delta{\bf H}_{\rm I}|n\rangle_d=O(\lambda^8 M^{-7})\per
\en
Consequently, at this order the theories described by
the Lagrangian~(\ref{eq:2-23}) and the FWT Lagrangian, yield the same
bound-state energy. We do not intend to construct  a proof that this
holds to any order.

\setcounter{equation}{0}
\section{Including the dynamical light field}
\label{sec:dynamical}

In section~\ref{sec:NRex}, we have set up a non-relativistic framework
that allows one to calculate the energy-levels of a particle moving in a given
external field. We wish to extend this construction to the model
discussed in section~\ref{sec:rel}, and  to consider
radiative  corrections
to the energy-levels, generated by the Yukawa interaction with the light
field.
In order to set up a non-relativistic framework for the calculation of
the energy shift, one has to specify the Lagrangian which will be used in
these calculations, and the rules that are used to obtain
observable quantities.
Indeed, the
perturbative expansion will be  carried out not only in the
coupling constant, but also in   energies and momenta,
because one needs to know the amplitude just in the vicinity of the
threshold. In such a multiple expansion, the question of ordering
arises - one has to specify the relative magnitudes of the expansion
parameters. This then determines what terms in the expansion should be
grouped together. A particular ordering is called {\em power counting
scheme}. The calculational framework
must then ensure that the ordering is not destroyed by loop
corrections - this problem goes under
the name ``validity of power counting''.

In this section, we discuss scattering amplitudes
at tree level and identify the problems that show up in loop
calculations.
 In order to keep the arguments simple, we put $\lambda=0$ in this
and in the following section.

\subsection{Tree graphs}
In a first step, we construct a non-relativistic Lagrangian that
reproduces the tree graphs generated by the relativistic Lagrangian
 Eq. (\ref{eq41}). We will restrict the discussion to
the one heavy particle sector, and first consider scattering  matrix elements
 in the relativistic theory,
\eq\label{eq51}
\langle {p},{q_{n+1}},\ldots, { q_m}\, \mbox{out}\, |
{k},{q_1},\ldots {q_{n}}\,\mbox{in}\rangle=
\langle \ldots \mbox{in}\,|\ldots\mbox{in}\rangle&+&
i(2\pi)^4\delta^4(P_f-P_i)T(p,k;q_1,\ldots,q_m)
\, ;
\nonumber\\[2mm]
&&\, m\ge 2\, , n\ge 1\per
\en
At tree level, the matrix element $T$ depicted in Fig.~\ref{scat-ell},
is a sum of products of free
propagators - each term is of the form

\eq\label{eq52}
e^m\prod_{j=1}^{m-1}\frac{1}{M^2-P_j^2}\,,
\en
with
\eq
P_j^0&=&k^0+\sum_{\nu=1}^{m}c_{\nu,j} q_\nu^0\, , \,
{\bf{P}}_j={\bf k}+\sum_{\nu=1}^{m}c_{\nu,j}{\bf q}_\nu\, ;
 \, c_{\nu,j}=\pm 1,0\per
\en
The low-energy expansion is now defined as follows. We decompose each
propagator in a singular and a regular part,
\eq\label{eq54}
\frac{1}{M^2-P_j^2}=\left(\frac{1}{\omega_j-P_j^0}
+\frac{1}{\omega_j+P_j^0}\right)\frac{1}{2\omega_j}\, ; \quad\quad
\omega_j=(M^2+{\bf P}_j^2)^{1/2}\, ,
\en
write in the first part
\eq
\omega_j=M+\frac{{\bf P}_j^2}{2M}+\triangle \omega_j=\bar{\omega}_j+\triangle
\omega_j\co
\en
and note that $\triangle \omega_j$ is much smaller than the difference
$\bar{\omega}_j-P_j^0$, because
\eq
\triangle \omega_j=O({\bf P}_j^4)\scs\quad\quad
\bar{\omega}_j-P_j^0=O(q_j^0)\, ,
\en
everywhere in momentum space except the specific configurations
of external momenta for which the difference $\bar{\omega}_j-P_j^0$
is of order ${\bf P}_j^4$ (these configurations have zero
measure in momentum space).
Therefore, one may expand in the first term in $\triangle
\omega_j$. Furthermore, in the second term and in the factor $1/\omega_j$,
 one expands in inverse powers of the heavy mass $M$. In this manner,
each of the propagators in each factor in the product (\ref{eq52})
becomes a series of terms of non-relativistic propagators. The leading
term reads
\eq
\frac{e^m}{(2M)^{m-1}}\prod_{j=1}^{m-1}\frac{1}{\bar{\omega}_j-P_j^0}\co
\en
whereas the next terms contain additional powers
$\triangle \omega_j/(\bar{\omega}_j-P_j^0)$.
 There are also terms
that do not contain any propagators - these originate from the second
term in Eq.~(\ref{eq54}). Finally, we also expand the quantities
$\triangle \omega_j$ in inverse powers of the heavy mass. As an example of
this procedure, we consider Compton scattering, which amounts to
$m=2,n=1$ in (\ref{eq51}). The intermediate particle
in the $s$-channel graph has four momentum
\eq
P_1^\mu=(k^0+q_1^0,{\bf k}+{\bf q}_1)\co
\en
and the matrix element becomes
\eq
e^2T_s =\frac{e^2}{2\omega_1}\left[\frac{1}{\omega_1-k^0-q_1^0}
+\frac{1}{\omega_1+k^0+q_1^0}\right]\scs \omega_1=(M^2+{\bf P}_1^2)^{1/2}\per
\en
Expanding in the manner described above, we obtain an infinite
series of terms,
\eq
T_s=\frac{1}{2M}\frac{1}{\bar{\omega}_1-k^0-q_1^0}
-\frac{{\bf P}_1^2}{4M^3}\frac{1}{\bar{\omega}_1-k^0-q_1^0}
+\frac{{\bf P}_1^4}{16M^4}\frac{1}{(\bar{\omega}_1-k^0-q_1^0)^2}
+\frac{1}{4M^2}+\cdots\per
\en
 These contributions have the following  representations as Feynman graphs:
non-relativistic propagation, vertex correction, mass insertion,
contact term, etc. The contribution in the $u$-channel may be treated
in an analogous way. In this manner, one generates a well-defined
series of terms, that may be ordered in inverse powers of the heavy
mass $M$.

The next step consists in the construction of a non-relativistic
Quantum Field Theory
that generates exactly this series for any tree graph in the one heavy
particle sector. We start from
the observation that the tree graphs are generated by the classical
action, evaluated at the solution to the classical EOM.
Retaining only the terms that correspond to the sector with one
heavy particle, the corresponding
non-relativistic Lagrangian is closely related to the one discussed in
section~\ref{sec:NRex} for the motion in the presence of the external field,
\eq
{\cal L}=\frac{1}{2}\,\,\partial_\mu \ell\partial^\mu \ell+
h^\dagger{\cal D} h\co\quad\quad
{\cal D}=D_+ + ed\ell d(1+eD_-^{-1}d\ell d)^{-1}\per
\en
The differential operators  $D_{\pm}$ and $d$ are defined in
section~\ref{sec:NRex}.
In order to arrive at a local Lagrangian, one again expands these
differential operators in inverse powers of $M$ - the result is given by
Eq.~(\ref{A2}) in Appendix~\ref{app:EOM}, with
$\lambda\phi\rightarrow e\ell$. This Lagrangian is
analogous to the effective Lagrangian (\ref{eq:2-23}), except that now,
the field $\ell$ is time dependent. As a result, some of the operators
contain time derivatives of the light
field $\ell$. Higher order time derivatives  can be eliminated
in a standard  manner, see
Appendix~\ref{app:EOM}. The final result  for the effective
Lagrangian  is of the form
\eq  \label{Lagrangian-ell}
{\cal L}_{NR}&=&
h^\dagger D_0 h + \frac{1}{2}\partial_\mu \ell\partial^\mu \ell
 + \sum_{n=1}^{\infty}\frac{1}{(2M)^n}\,{\cal L}_n\, ,
\nonumber\\[2mm]
{\cal L}_1=e h^\dagger\ell h\, ,\quad\quad {\cal L}_2&=&0\, ,\quad\quad
{\cal L}_3=h^\dagger(\triangle^2+e(\ell\triangle + \triangle\ell)+e^2\ell^2)h
\, ,\quad\ldots\per
\en
The scattering matrix elements are obtained by evaluating - at tree
level - the connected Green functions
\eq
G_{NR}(x,y;z_1,\ldots,z_m)=i\langle0|Th(x)
h^\dagger(y)\ell(z_1)\ldots \ell(z_m)|0\rangle_c\, .
\en
The perturbation theory is performed as follows. First, one sums up
all mass insertions in the external heavy lines. The Fourier transform
of $G_{NR}$ then develops poles, and one
ends up with a reduction formula for the scattering amplitude that is
similar to Eq.~(\ref{eq:2-26}), except
that there are also poles corresponding to the light fields.
This procedure generates the low-energy expansion of the relativistic
scattering amplitude discussed above.

\subsection{Loops}
The evaluation of tree diagrams is performed by expanding in inverse
powers of the heavy mass $M$. Each term in the Lagrangian contributes
in a definite order - the expansion makes perfect sense. By
calculating loops, one again needs  an expansion where
the contributions from higher dimensional operators are suppressed in
a well defined manner - otherwise, the framework is not useful.
This problem has obtained an extensive coverage in the literature, in the
context of different physical
problems~\cite{Manohar,Griesshammer,HBChPT,Becher}.
We first illustrate the problem in the case of the correlator of
two heavy fields. In the following section, we show  that the above
Lagrangian does lead to  a
consistent framework, provided that the rules to evaluate the Feynman
diagrams are properly adapted.

In order to avoid inessential technical complications, we  consider
here the
two-point function at vanishing 3-momentum ${\bf p}$ in $D$ dimensions,
\eq
G_{NR}(\sigma)=i\int d^Dxe^{ip^0x^0}\langle
0|Th(x)h^\dagger(0)|0\rangle
=\frac{1}{M-p^0-\Sigma_{NR}(\sigma)}\sem \sigma=2(M-p^0)/M\per
\en
Here, we have introduced the dimensionless variable $\sigma$ for
 later convenience, and use  the
quantity $\Sigma_{NR}$ to collect  the contributions
from the loops, as usual. The dependence of the propagator and of the
self-energy part on the variables $e^2$ and $M$ is suppressed to
ease notation. In the following, we concentrate on the
evaluation of $\Sigma_{NR}$ at lowest non-trivial order $e^2$. Some of
the diagrams that contribute at this order are depicted in
Fig.~\ref{figselfnr}. The contribution from diagram Fig.~\ref{figselfnr}A
reads
\eq\label{eq515}
\Sigma_{NR}^{A}=\frac{e^2}{4M^2}\frac{1}{i(2\pi)^D}\int\frac{d^Dl}{-l^2}
\frac{1}{M-p^0+l^0+\frac{{\bf l}^2}{2M}}\per
\en
Without performing any explicit calculation, we can make the following
observations. First, the integral is ultraviolet divergent - one has to add
counterterms to the original Lagrangian to render the propagator
finite. The divergence is not suppressed in the large $M$ expansion -
the counterterm required contributes to the mass term.
 Further, expanding the integrand in powers of $\sigma$, it
is seen that there will be terms analytic at $\sigma=0$, as well as
a non-analytic term  $\sim \sigma
\ln{\sigma}$, reflecting the fact that the propagator generates a
branch point at $p^0=M$. The contributions from
the diagrams Fig.~\ref{figselfnr}B~+~\ref{figselfnr}C are
given by the integral
\eq\label{eq516}
&&\Sigma^{B+C}_{NR}=-\frac{e^2}{8M^4}\frac{1}{i(2\pi)^D}
\int\frac{d^Dl}{-l^2}\,\,{\bf l}^2\,\,
\frac{1}{M-p^0+l^0+\frac{{\bf l}^2}{2M}}\per
\en
Again, the divergence of this integral is not suppressed in
the momentum expansion - the mass term is affected also here.
 Further, there are  contributions  analytic in $\sigma$,
 as well as a non-analytic term
 $\sim \sigma^3\ln{\sigma}$. We conclude from these two examples that
perturbation theory, based on the Lagrangian (\ref{Lagrangian-ell}) and using
standard dimensional regularization, generates a complicated series of
terms: in general,  all operators in the effective Lagrangian
will  contribute e.g. to the mass. This feature of the standard
expansion is very similar to the problems already discussed in application
to different physical problems \cite{Manohar,Griesshammer,GSS,HBChPT,Becher}.

The reason for the breakdown of the counting rules lies in the inconsistency
of the procedure:
the non-relativistic effective theory we have constructed
is supposed to be valid at very small momenta. On the other hand,
 using dimensional
regularization, we have extended the momentum integration to infinity.
In the presence of the hard mass scale $M$, the propagators are no longer
homogeneous functions of momenta, and  the procedure fails to
suppress contributions from higher dimensional operators.
The problem is clearly caused by the contributions from large
momenta in the integral. Indeed, at very small momenta, the term
${\bf l}^2/(2M)$ in the heavy particle propagator may be dropped -
 then,the integrand is homogeneous
in the momenta, and the contributions from higher dimensional
operators are suppressed, being proportional to $\sigma^{D-3}$
($\sigma^{D-1}$) for the graph depicted in Fig.~\ref{figselfnr}A
(Fig.~\ref{figselfnr}B, Fig.~\ref{figselfnr}C).
 On the other hand, the high energy contributions generate a
contribution that is analytic at $\sigma=0$.
 One needs  a systematic procedure
to subtract  these unphysical high-mom\-en\-tum contributions from the
loop diagrams.
 The problem has been discussed in
the recent literature \cite{QED,Manohar,Beneke,Griesshammer}.
The prescriptions offered  reduce to a set of rules which
 are  applied to the integrands,
 before the integration is carried out. We illustrate it in the
following section.

\section{The two-point function}
\label{sec:twopoint}
\setcounter{equation}{0}
In order to illustrate the method that allows one to subtract the
 uninteresting polynomial part, we consider   here in detail the two-point
 function of the heavy field at order $e^2$, in the absence of the
external field.

\subsection{The two-point function at {\bf p}=0}
We  first consider again the case where the three momentum is set to
zero, and start
with the  expression (\ref{eq515}). Combining the denominators by
use of the Feynman-Schwinger type formula
\eq
\frac{1}{ab}=\int_0^\infty dy\frac{1}{(a+by)^2},
\en
one may  perform all integrations, with the result
\eq\label{eq519}
\Sigma_{NR}^{A}&=&-\frac{e^2}{16\pi^2M}\left\{
 \Phi_1\ln\sigma +\Phi_2\right\}+O(D-4)\co\nonumber\\[2mm]
\Phi_1&=&1-\frac{1}{W}\scs\quad
\Phi_2=2[8\pi^2\bar L_M+\ln 2-1+\frac{1}{W}\ln{(1+W)}]\co\quad
W=\sqrt{1-\sigma}\per
\en
In order to remove the pole in  $\Phi_2$, a mass counterterm is
needed. In order to keep the
position of the pole at $p^0=M$, the finite part of the counterterm
has to be appropriately tuned. In addition, it is explicitly seen
 that the self-energy develops a  logarithmic branch point at
$p^0=M$. This singularity is suppressed, in the sense that it is
proportional to $\sigma$.

 We now compare this result with
what one obtains through the multipole
expansion \cite{QED,Manohar,Beneke,Griesshammer}.
In the present case, this method amounts to expand the integrand in
(\ref{eq515}) in the three-momentum of the massless particle,
according to the counting
\eq\label{eq517}
p^0-M = O(v^2),  \,  |\vec{p}|=O(v)\, , \quad\quad
l^0 = O(v^2),  \,  |\vec{l}|=O(v^2)\, ,
\en
where $v$ is a small parameter. Proceeding in this manner, an infinite
series of terms is generated out of the quantity $\Sigma_{NR}^{A}$,
\eq\label{series}
\hat{\Sigma}_{NR}^{A}
&=&\sum_{n=0}^\infty\Sigma_n(\sigma)\co\nonumber\\
\Sigma_n&=&(-1)^n\frac{e^2}{4M^2}\frac{1}{i(2\pi)^D}
\int\frac{d^Dl}{-l^2}\,\,\left(\frac{{\bf l}^2}{2M}\right)^n\,\,
\frac{1}{(M-p^0+l^0)^{n+1}}\per
\en
The hat on $\hat{\Sigma}_{NR}^A$ indicates that the result of this
manipulation is not identical to the previous expression
$\Sigma_{NR}^A$. It is seen that  $\Sigma_n$ is of order
$v^{2D+2n-6}$
near threshold. Let us consider the first term in the series,
\eq\label{eqfirst}
\Sigma_0=\frac{e^2}{4M^2}\frac{\Gamma(D-2)\Gamma(3-D)}
{(4\pi)^{\frac{D-1}{2}}\Gamma(\frac{D-1}{2})}(M-p^0)^{D-3}\per
\en
In contrast to the case considered before, this term is now suppressed
near threshold, being of order $v^2$ at $D\rightarrow 4$.
The mass is unaffected by the loop correction, and the
counterterm needed to cancel the ultraviolet divergence is also of
order $v^2$, provided that we count the differential operator
$i\partial_t-M+\frac{\triangle}{2M}$ as a quantity of order $v^2$.
Furthermore, the non-analytic part in this first term,
\eq
\Sigma_0=\frac{e^2}{32\pi^2M}\sigma\ln\sigma+\cdots\co
\en
exactly agrees with the non-analytic part in the original expression
(\ref{eq519}). This feature is true to all orders in the $v^2$
expansion (\ref{series}). Indeed, the integrations can be done
 in each term, by integrating
first over $l^0$ in the lower half plane.
In the limit $D\rightarrow 4$, the summation can be explicitly
performed, with the result
\eq\label{SE-mpole}
\hat{\Sigma}_{NR}^A&=&
-\frac{e^2}{16\pi^2M}\left\{
\Omega_1\ln\sigma +\Omega_2\right\}\co\nonumber\\
\Omega_1&=&\Phi_1,\quad\quad
\Omega_2=\left(1-\frac{1}{W}\right)\left\{8\pi^2\bar L_M+\ln 2
-1 -\ln(1+W)\right\}\per
\en
It is seen that indeed, the logarithmic part in $\Sigma_{NR}^A$ is exactly
reproduced, whereas the second contribution $\Phi_2$, analytic at
threshold,  is modified, as a result of which $\Sigma_n$ is of the
order expected from counting powers of $v$ in the integrand.
[The summation of the series in
Eq.~(\ref{series}) has been carried out for illustrative purpose only -
in order to demonstrate that the infrared-singular part of the
multipole-expanded self-energy diagram indeed reproduces the exact result
given by Eq.~(\ref{eq519}) in all orders in the expansion in $\sigma$.
In practice such a summation is never used - instead, only a finite number
of terms is retained that contribute to the matching condition in a given
accuracy.]

The self-energy (\ref{SE-mpole}) has,  aside from  the branch point at
$p^0=M$,  in addition a branch point at $p^0=M/2$ on the first
sheet in the
complex $p^0$ plane. This additional branch point, however, lies outside
the region $p^0\sim M$ where both the non-relativistic approach and the
multipole expansion are supposed to be meaningful. The close analogy
with the ``infrared regularization'' of Feynman integrals in Baryon
ChPT~\cite{Becher} is evident. Further, it is seen
from Eq.~(\ref{SE-mpole}) that the ultraviolet-divergent term is no
longer polynomial in momenta after the summation -
the renormalization procedure can be carried out only order by
order in the $v^2$ expansion.

 We conclude that in this example, the two expressions $\Sigma_{NR}^A$
and $\hat{\Sigma}_{NR}^A$ differ by
 a power series in $\sigma$. Next, we
investigate the contributions from the graphs
Fig.~\ref{figselfnr}B~+~\ref{figselfnr}C.
 Proceeding analogously, it is seen
 that these start to contribute now at order $\sigma^{D-1}$. We have not
carried out the summation of the multipole expanded terms in this
case. However, we have checked that the leading term in the logarithmic
singularity again agrees with the one in the original expression
$\Sigma_{NR}^{B+C}$ in Eq.~(\ref{eq516}).

We see  that the multipole expansion
eliminates the unwanted terms that do not respect power counting.
 In addition, it is clear from the procedure that contributions generated
by the higher order terms in the effective Lagrangian will  be
suppressed in the momentum expansion in an analogous manner.
At each order in the expansion, the ultraviolet divergences may be
removed by adding a local term to the effective Lagrangian. These
terms again are suppressed by inverse powers of $M$, as a result of
which one has a perfectly consistent procedure. The self-energy
$\hat \Sigma_{NR}$ becomes in the multipole expansion
\eq
{\hat{\Sigma}}_{NR}&
=&\sum_{n=1}^{\infty}\sigma^n\left\{c_n\ln{\sigma}
+d_n\right\}+O(D-4)\per
\en
The terms of order $\sigma^n$ are generated by a {\em finite} number of
diagrams. Further, the branch point in the
propagator  stays at $M$ at each order in this expansion, there is no
need to adjust the mass term in the Lagrangian for this to achieve.

\subsection{The two-point function at ${\bf p}\neq 0$.}
So far, we have considered the propagator at zero three momentum.
Let us now consider the general case,
\eq\label{eq524}
G_{NR}(p)=i\int d^Dx e^{ipx}\langle 0|Th(x)h^\dagger(0)|0\rangle\per
\en
As before, we consider the corresponding self-energy at order $e^2$.
The complication that arises for non-vanishing three momentum is the
following. The contributions from the mass insertions
$h^\dagger(\Delta/M^2)^n h$ generate singular contributions near
$p^0=M+{\bf p}^2/(2M)$ for small three momenta~(see Fig.~\ref{insertions}),
of the form
\eq\label{eq525}
\mbox{const}\times e^2{\bf p}^{2n}/(M+\frac{{\bf p}^2}{2M}-p^0)^m\per
\en
The origin of  this singularity becomes very clear in the
 multipole expansion. Indeed,
summing up the most singular parts of these divergences, one finds
that the self-energy behaves as
\eq\label{eq526}
\hat{\Sigma}_{NR}(z,{\bf p}^2)
=\frac{e^2}{16\pi^2M}z\ln{z}+\cdots\scs
\quad\quad z=(\omega_{\bf p}-p^0)/M\co
\en
 near the threshold $z=0$. Whereas the lowest
order diagram Fig.~\ref{figselfnr}A generates a
branch point at $p^0=M+{\bf p}^2/(2M)$, the mass insertions move this
to $p^0=\omega_{\bf p}$, as it is required from the fact that
the theory should describe the non-relativistic limit of the
relativistic theory \cite{Gall}.
This is analogous to the free propagator, that has, in the absence of
any interaction, a pole at $p^0=M+{\bf p}^2/2M$ - only after an
infinite re-summation of mass insertions does this change to the
required form $1/(\omega_{\bf p}-p^0)$.

The
calculation of the non-relativistic propagator may now be performed
as follows. All calculations are done using the multipole expansion,
counting $M+{\bf p}^2/(2M)-p^0$ and ${\bf p}^2$ as quantities of order
$v^2$. At each order in this expansion, the
counterterms required may be chosen such that $\hat\Sigma_{NR}$
remains finite at $D\rightarrow 4$,
\eq\label{eq528}
\hat\Sigma_{NR}(z,{\bf p}^2)&=&
\frac{e^2}{16\pi^2M}\sum_{n=1}^{\infty}z^n\left\{c_n' z^{D-4}
+d_n'\right\}\nonumber\\&=&
\frac{e^2}{16\pi^2M}\sum_{n=1}^{\infty}z^n\left\{\bar{c}_n\ln{z}
+\bar{d}_n\right\}+O(D-4)\co
\en
where the coefficients $c_n',~d_n'$ and $\bar{c}_n,~\bar{d}_n$
are functions of the three-momentum, $c_n'=c_n'({\bf p}^2)$, etc.
This procedure generates a well defined series at each order in the
expansion, and  each term in the series (\ref{eq528}) is again
generated by a finite number of diagrams.
The behavior of the two-point function near the pole at
$p^0=\omega_{\bf p}$ is given by
\eq\label{pole-behavior}
\hat G_{NR}(p)\equiv
\frac{1}{\omega_{\bf p}-p^0-\hat\Sigma_{NR}(z,{\bf p}^2)}\rightarrow
\frac{Z_{NR}}{\omega_{\bf p}-p^0}+ O(z^{D-5})\co
\en
with
\eq\label{Z-nonrel}
Z_{NR}=1+\frac{e^2}{16\pi^2M^2}\, d_1'+O(e^4)\per
\en

Equation~(\ref{pole-behavior}) may be used to define the mass in the
non-relativistic theory - it is the parameter $M$ with the property that the
two-point function  has a pole at $p^0=\sqrt{M^2+{\bf p}^2}$.
For a detailed discussion of this point, see Ref.~\cite{Gall}.

For later use, we give the explicit expression of the multipole-expanded
self-energy in the non-relativistic theory at order $e^2v^2$. At this order,
only the graph shown in Fig.~\ref{figselfnr}A contributes. One should add the
counterterm that is needed to cancel the ultraviolet divergence - the
corresponding Lagrangian is given by Eq.~(\ref{Lct}). The self-energy is given
by
\eq\label{SE-v2}
\hat\Sigma_{NR}&=&
\frac{e^2}{4M^2}\,\,\frac{(Mz)^{D-3}}{(4\pi)^{\frac{D-1}{2}}}
\,\,\frac{\Gamma(D-2)\Gamma(3-D)}{\Gamma(\frac{D-1}{2})}-
\frac{e^2f_1z}{4M}
\nonumber\\[2mm]
&=&\frac{e^2z}{16\pi^2M}\,\biggl\{\,
\ln 2z+\ln\frac{M}{\mu}-1-f_1^r(\mu)\biggr\}+O(D-4)\, ,
\en
where at $O(v^2)$, $Mz=M+{\bf p}^2/(2M)-p^0$, and
\eq\label{f_1}
f_1=\frac{1}{4\pi^2}\,(\bar L(\mu)+f_1^r(\mu))+O(e^2)\, .
\en

{}From Eq.~(\ref{SE-v2}), one can read off the expression for the
wave function renormalization factor,
\eq\label{Zfact}
Z_{NR}=1-\frac{e^2f_1}{4M^2}+O(e^2v^2,e^4).
\en

\subsection{Comparison with the relativistic theory}

It is instructive to compare at this stage these results with the
two-point function of the relativistic theory considered
in section~\ref{sec:rel}.  In the absence of the external field, one has
\eq\label{SE-D}
i\int d^D x\, e^{ipx}\,\langle 0|TH(x)H^\dagger(0)|0\rangle=
\frac{1}{M^2-p^2-\Sigma(p^2)}.
\en
The relevant contributions at order $e^2$ are displayed in
 Fig.~\ref{UV}A,~\ref{UV}D. Adding the counterterms, we find
\eq\label{SE-rel-1}
\Sigma(p^2)&=&\frac{e^2}{(4\pi)^{D/2}}\,\Gamma(2-\frac{D}{2})\,
\int_0^1 dx\, [xM^2-x(1-x)p^2]^{D/2-2}-\delta M^2\per
\en
Performing the limit $D\rightarrow 4$ before going to the mass-shell, we
obtain
\eq
\label{SE-rel-2}
\Sigma(p^2)=
\frac{e^2}{16\pi^2}\,\frac{M^2-p^2}{p^2}\,\ln\frac{M^2-p^2}{M^2}
+O(D-4)\, .
\en
Now, changing the sequence of the limiting procedures and
differentiating $\Sigma$ at $p^2=M^2$ and $D>4$,
one obtains the wave function renormalization constant $Z$,
\eq\label{Z-rel}
Z=1-\frac{e^2}{2M^2}\,\bar L_M+\frac{e^2}{16\pi^2M^2}+O(e^4)\per
\en

By comparing the leading logarithmic singularity in $\Sigma$
with the one in $\hat\Sigma_{NR}^A$ given by Eq.~(\ref{eq526}),
it is seen that they are identical, up to a factor
$1/2\omega_{\bf p}$. Based on the results in this section, we expect
that  higher order non-analytic terms in the relativistic
self-energy will be reproduced by
the multipole expansion as well. The polynomial part can then be adapted by a
suitable choice of the counterterms, as a result of which one obtains
\eq\label{eq527}
\hat{\Sigma}_{NR}(z,{\bf p}^2)=\frac{1}{2\omega_{\bf p}}
\,\,\Sigma(p^2)\scs p^2\simeq M^2\co
\en
order by order in the $v$ expansion.
We refer the interested reader to Ref. \cite{Gall} for a discussion of
this relation beyond one-loop accuracy.

\setcounter{equation}{0}

\section{Power counting  and matching}
\label{sec:matching}
We now introduce the external field $\phi$ again, and consider
 radiative  corrections  of order $e^2$ (one loop) to the scattering
 matrix elements discussed in
section III. We  restrict the discussion to Green functions that are
relevant for  bound state calculations considered later on -
i.e., we consider the one heavy particle sector in Fock space, with no
 incoming or outgoing light particles.

\subsection{The Lagrangian, the reduction formula and  power counting}
To generate the tree graphs, one  replaces
the external field $\lambda\phi$ in
 the Lagrangian given in Eq.~(\ref{A2}),
$\lambda\phi\rightarrow\lambda\phi+e\ell$, and adds the kinetic term for the
light field.
 Because we do not consider incoming/outgoing light particles,
 one needs to retain in the interaction Lagrangian only
terms that are linear in the  field $\ell$.
 In the presence of
radiative corrections, as we have already seen, one has to equip
this Lagrangian with
 counterterms that render the theory ultraviolet finite,
\eq\label{eq61}
{\cal L}_{NR}&=&
h^\dagger \biggl(\, i\partial_t-M+\frac{\triangle}{2M}\biggr)h
+ \frac{1}{2}\,(\partial_\mu\ell)^2
+ \sum_{n=1}^\infty \frac{1}{(2M)^n}\,{\cal L}_{n}
+ {\cal L}_{c.t.}\scs
\nonumber\\[2mm]
  {\cal L}_{1} &=& h^\dagger(g_1\,\lambda\phi+g_2 e\ell)h
\scs\quad\quad {\cal L}_{2}= 0\scs
 \nonumber\\[2mm]
  {\cal L}_{3} &=&  h^\dagger(\triangle^2+g_3\lambda(\phi\triangle
+ \triangle\phi) + g_4\lambda^2\phi^2
 +   g_5e(\ell\triangle+\triangle\ell)+2g_6e\lambda\phi\ell)h\scs
\nonumber\\[2mm]
\label{Lct}
{\cal L}_{c.t.}&=&\frac{e^2f_1}{4M^2}\,\,h^\dagger\,
\biggl(i\partial_t-M+\frac{\triangle}{2M}\biggr)h
+\cdots\per
\en
The coefficients
$g_i$ appearing in the Lagrangian are power series in the variable
$e^2/M^2$,  $g_i=1+O(e^2)$, and $f_1$ was determined in~(\ref{f_1})
at leading order in $e^2$. Scattering matrix elements
are calculated by use of the reduction formula Eq.~(\ref{eq:2-26})
except that now - in analogy to the relativistic framework -
the external legs of the heavy
particles are equipped with  wave function renormalization factors
$Z_{NR}^{1/2}$, and the radiative corrections to the external legs are
omitted.
 The matrix elements are then expanded in
powers of the coupling constants $\lambda$ and $e$,
and of the external momenta.  We define
the power counting scheme - that specifies how the momentum expansion
is done - by associating the following powers of $v$ to the
 derivatives in (\ref{eq61}),
\eq\label{eq62a}
i\partial_t-M+\frac{\triangle}{2M}=O(v^2)\scs
\triangle =O(v^2)\per
\en
[Note that one may arrange the derivatives in such a manner that the
Laplacian never acts on the light field.]
In addition, it is understood that the loops are evaluated by use of
the multipole expansion considered above: the loop momenta of the light
particle are counted as quantities of order $v^2$, and the propagators
of the heavy field are expanded accordingly.

One could now set up a formal power counting for generic graphs containing
generic vertices from (\ref{eq61}). However, we find it more instructive
to consider instead two examples, the self-energy discussed in
the previous section, and radiative corrections to scattering in the
external field at lowest order.

We start the discussion  with the self-energy - a few graphs are displayed in
Fig.~\ref{figselfnr}. According to (\ref{eq62a}), we count the heavy
(light) propagator as a quantity of order $1/v^2$ ($1/v^4$), as a
result of which the graph Fig.~\ref{figselfnr}A is counted as a term of order
 $v^2$.
 As we have discussed  in the previous section, the multipole
 expansion transforms the relevant integral into an infinite string of terms,
starting at $O(v^2)$.
 In addition, this graph requires, in the multipole expansion,
a string of counterterms that start with the one proportional to $f_1$
in (\ref{eq61}). This is indeed a term of order $v^2$ according to the
above counting. The vertices
in Fig.~\ref{figselfnr}B,~\ref{figselfnr}C stem from ${\cal L}_3$ and generate
 therefore additional factors of at least the order  $v^2$.
 The corresponding string of counterterms
 starts at  order $v^4$, etc. We now turn to scattering in the
external field at order $\lambda$.
 Some of the relevant graphs are displayed in
 Fig.~\ref{figvertex-NR}. The tree graph Fig.~\ref{figvertex-NR}A  has been
evaluated in
(\ref{eq:2-30}).
 The radiative  corrections in Figs.~\ref{figvertex-NR}B and
\ref{figvertex-NR}C
 amount to multiply the tree graph with
 the residue $Z_{NR}$ of the propagator -  this contribution  starts at order
$e^2\lambda v^0$. The vertex correction Fig.~\ref{figvertex-NR}D starts at the
same order (we
discuss below the manner in which one tames the
ultraviolet and infrared divergences hidden here). Finally, the graphs
Fig.~\ref{figvertex-NR}E and Fig.~\ref{figvertex-NR}F contain one vertex from
${\cal L}_3$, proportional to $g_6$.
The leading term of this contribution is suppressed by $v^2$, because
it contains one heavy propagator only.

These two examples show two things: First, one
 associates  to any tree or one-loop graph a factor  $v^m$ according
 to the rules spelled out above. The multipole
 expansion turns the loop contribution into an infinite string of
 terms, such  that the amplitude
 starts at $O(v^m)$. Second, contributions
 from higher order operators ${\cal L}_n$ are suppressed in the
 following sense. Consider loop corrections to a matrix element at
 some fixed order $\lambda^n$, and compare the corrections due to
 a loop generated by ${\cal L}_1\times {\cal L}_p$  with the one
generated by ${\cal L}_1\times
 {\cal L}_q$, with $p>q$. Then, the leading contribution from the former is
 suppressed with respect to the one of the latter, and the relevant string of
 counterterms needed the renormalize the former starts at higher order
 than the one for the latter: although individual graphs do
 not contribute at a single order, the starting point of the expansion
 is hierarchically ordered.

\subsection{Matching: The vertex function at order $e^2$}

In order to determine the coefficients $g_i$ in (\ref{eq61}),
one evaluates the amplitudes at a fixed order in $\lambda,e$
 and requires that the results be the same in the
relativistic and in the non-relativistic theory at a given order in
$v$. As the effect of the
operators of higher dimensions are suppressed in the multipole
expansion, only a finite number of coefficients contributes at a fixed
order $v$. In the following section, we will show that
we need only the coefficient $g_1$ at order $e^2$ for our purpose.
It suffices to evaluate a matrix element that fixes this term, and we
choose scattering of the heavy particle in the external field $\phi$
at order $e^2\lambda$.

The relevant diagrams in the relativistic theory
are displayed in Fig.~\ref{figvertex} (with counterterms omitted),
while those of the non-relativistic
 scattering matrix $T_{NR}$ are shown in Fig.~\ref{figvertex-NR}.
 As we are considering a
massless theory, we use dimensional regularization to also regularize
the infrared divergences.

The matching condition
of relativistic and non-relativistic scattering amplitudes reads
\eq\label{eq63}
T_{NR}({\bf p},{\bf k})=T({\bf p},{\bf k})\, .
\en
At order $\lambda$, these scattering amplitudes can be written as
\eq\label{eq61a}
T({\bf p},{\bf k})&=&
\lambda \phi({\bf p}-{\bf k})\tilde T({\bf p},{\bf k})\, ,
\nonumber\\[2mm]
T_{NR}({\bf p},{\bf k})&=&
\lambda \phi({\bf p}-{\bf k})\tilde T_{NR}({\bf p},{\bf k})\, .
\en
At $D\neq 4$, both $T$ and $T_{NR}$ can be expanded in the momenta.
Since we only need to determine the coefficient $g_1$, it suffices to work
at vanishing momenta, and to require that
\eq\label{eq64}
\tilde T_{NR}({\bf 0},{\bf 0})=\tilde T({\bf 0},{\bf 0})\per
\en
We start the calculation with the relativistic amplitude.
The contribution from diagram Fig.~\ref{figvertex}E has been
evaluated in section~\ref{sec:rel}.
The diagram Fig.~\ref{figvertex}D is ultraviolet finite and easy to
calculate at zero
three-momenta in $D$ dimensions. Adding the
contribution from the wave function renormalization constant $Z$
gives
\eq\label{eq65}
\tilde T({\bf 0},{\bf 0})=1+
\frac{5e^2}{96\pi^2M^2}+O(e^4)\per
\en
As expected, the infrared divergences that are  present  in the
 residue $Z$  and in the vertex function Fig.~\ref{figvertex}D,
cancel at threshold. Indeed, infrared finite
cross sections are obtained by adding diagrams from Bremsstrahlung.
 At threshold, the phase space for these processes vanishes  -
therefore, the amplitude must be infrared finite itself.

Next we calculate the amplitude in the non-relativistic theory.
The vacuum polarization diagram is absent in this case,
while the contributions
 from Fig.~\ref{figvertex-NR}E,~\ref{figvertex-NR}F are suppressed
in the $v$ expansion.
The contribution from diagram Fig.~\ref{figvertex-NR}D is ultraviolet
divergent in the multipole expansion. In order to separate this
divergence from the one in the infrared region, we proceed as
follows.  We first consider the amplitude off-shell,
\eq\label{eq66}
\Gamma_{NR}(p^0)=
\frac{e^2}{4M^2i(2\pi)^D}\int\frac{d^D l}{-l^2}
\frac{1}{(M-p^0+l^0)^2}\, .
\en
Evaluating this expression in $D$ dimensions gives
\eq\label{eq67}
\Gamma_{NR}(p^0)
=\frac{e^2}{M^2}\,
\frac{\Gamma(D-2)\Gamma(4-D)}
{4(4\pi)^{\frac{D-1}{2}}\Gamma(\frac{D-1}{2})}\,(M-p^0)^{D-4}\, .
\en
Now, we  eliminate its divergence at $D=4$ by using
\eq\label{eq69}
g_1=1+\frac{e^2}{16\pi^2M^2}\,(\bar L(\mu)+g_1^r(\mu))+O(e^4)\co
\en
where $g_1^r(\mu)$ denotes the finite part in the coupling
constant $g_1$ at order $e^2$.
Finally, keeping $D>4$, multiplying with $Z_{NR}$ the
lowest-order contribution displayed in Fig.~\ref{figvertex-NR}A
and summing up all contributions, we obtain
\eq\label{TNR}
\tilde T_{NR}({\bf 0},{\bf 0})=
1+\frac{e^2}{16\pi^2M^2}\,(g_1^r(\mu)-f_1^r(\mu))+O(e^4)\, .
\en
[Remark: under the following sequence
 of limiting procedures:
going to mass-shell $p^0\rightarrow M$
at  $D>4$ before performing the limit $D\rightarrow 4$, the
unrenormalized vertex function $\Gamma_{NR}$ vanishes.
It is only the polynomial part proportional to $g_1$, that survives in
this limit.]

The matching condition (\ref{eq63}) now determines the difference of the finite
parts of the counterterms,
\eq\label{eq68}
g_1^r(\mu)-f_1^r(\mu)=\frac{5}{6}+O(e^2)\, .
\en

\subsection{The Hamiltonian}

In order to calculate the energy-levels, we need to construct the
Hamiltonian pertaining to  the Lagrangian (\ref{eq61}). As is discussed in
Appendix~\ref{app:EOM}, we first re-scale the heavy field, such that the
kinetic term
has coefficient one. As we already mentioned and as is proven in the following
section, only the terms proportional to $g_{1,2}$ are needed in the
calculation of the energy-level at order $e^2\lambda^2$. Further, terms
of order $e^2$ in $g_2$ do not contribute.
Therefore, re-scaling the
heavy field amounts to the renormalization
\eq\label{eq460}
g_1\rightarrow g_1-\frac{e^2}{4M^2}\, f_1=1+\frac{5e^2}{96\pi^2M^2}\co
\en
as a result of which the Lagrangian becomes
\eq\label{eq461}
&&{\cal L}_{NR}=
h^\dagger \biggl(\, i\partial_t-M+\frac{\triangle}{2M}\biggr)h
+ \frac{1}{2}\,(\partial_\mu\ell)^2
+h^\dagger\biggl[\, \frac{e}{2M}\ell+\frac{\lambda}{2M}\biggl( 1+
\frac{5e^2}{96\pi^2M^2}\biggr)\,\phi\biggr]\,h +\cdots \, \per
\nonumber\\
&&
\en
The ellipsis denotes terms that are not needed in the following.

{\underline{Remark:}} This Lagrangian was obtained without specifying
 the couplings $f_1,g_1$ - only the
combination (\ref{eq460}) counts, which is fixed by the matching condition
(\ref{eq68}). The Lagrangian (\ref{eq461}) amounts to using
$f_1=0$, which leads  to $Z_{NR}=1$ at this order in the low-energy expansion.
As can be seen from the explicit expression (\ref{SE-v2}), the
derivative of the non-renormalized self-energy indeed vanishes at
$D>4$, which is  a particular manifestation  of so-called
 ``no-scale'' arguments.
 In this setting, the self-energy is not finite at $D=4$. However, the
 pole position of the two-point function is not affected, and
 $S$-matrix elements remain the same.\footnote{We are
under the impression that many of the contemporary
 bound-state calculations use implicitly an argument of this kind. On the
 other hand, we are not aware that the relation of this procedure to standard
 renormalization theory has been made explicit in the literature.}

Finally, we display  the Hamiltonian. Denoting by
 $\Pi_\ell$  the canonically conjugated momentum of
the light field, one has
\eq\label{eq462}
{\cal H}&=&{\cal H}_0+{\cal H}_C+{\cal H}_I\co
\en
with
\eq\label{eq462_1}
{\cal H}_0&=&
{h^\dagger\biggl( M-\frac{\triangle}{2M}\biggr)h
+\frac{1}{2}\,\Pi_\ell^2+\frac{1}{2}\,(\vec\nabla\ell)^2}\scs
\nonumber\\[2mm]
{\cal H}_C&=&-{h^\dagger\,\frac{\lambda\phi}{2M}\, h}\scs
{\cal H}_I=-{\frac{e}{2M}\,h^\dagger\ell h
-\frac{5e^2\lambda}{192\pi^2M^3}\,h^\dagger\phi h}\per
\en
We use in the last section these expressions  to calculate the energy-level
shift of the ground state.

\setcounter{equation}{0}
\section{The ground state}
\label{sec:groundstate}

The radiative shift of the  energy-levels is
calculated by use of Eq.~(\ref{PT}),
\eq\label{eq463}
\Delta E_n&=&\langle n|{\bf H}_{\rm I}|n\rangle+\sum_{m\neq n}
\frac{\langle n|{\bf H}_{\rm I}|m\rangle\langle m|{\bf H}_{\rm
I}|n\rangle}{E_n-E_m}
+\sum_{m,\gamma}
\frac{\langle n|{\bf H}_{\rm I}|m,\gamma\rangle
\langle m,\gamma|{\bf H}_{\rm I}|n\rangle}{E_n-E_m-E_\gamma}
\nonumber\\[2mm]
&-&\Delta E_n\sum_{m\neq n}
\frac{\langle n|{\bf H}_{\rm I}|m\rangle\langle m|{\bf H}_{\rm
    I}|n\rangle}{(E_n-E_m)^2}+\cdots\, ,
\en
where $\gamma$ denotes the light particle which is present in the sum over
intermediate states in the Fock space.
The Hamilton operator is constructed from the Lagrangian displayed
in Eqs.~(\ref{eq61}). The interaction
Hamiltonian is no longer diagonal and induces transitions between the Fock
space vectors with an arbitrary number of light particles/antiparticles. The
unperturbed wave functions in this expansion are the eigenfunctions of
${\bf H}_{\rm 0}+{\bf H}_{\rm C}$,
\eq\label{eq464}
({\bf H}_{\rm 0}+{\bf H}_{\rm C})|n\rangle=E_n|n\rangle,\quad
({\bf H}_{\rm 0}+{\bf H}_{\rm
  C})|n,\gamma\rangle=(E_n+E_\gamma)|n,\gamma\rangle\, .
\en
The Hamilton operator contains
an infinite string of operators. Only
a few of them, with the low mass dimension, contribute to the energy shift
in a given accuracy.
For the lowest-order term in (\ref{eq463}), or for
the sums with no light particle in the intermediate state,
one can  apply power-counting arguments:
Since the Coulomb wave functions do not depend on the heavy mass $M$, and the
energy denominators are homogeneous functions of $M$, one can count
the powers of $M$ in the energy shift once one knows
the explicit powers of $M$ in the couplings of various operators in
${\bf H}_{\rm I}$. Given more powers of $M$ in the denominator,
one needs additional powers of $\lambda$ in the numerator, in order
to compensate the extra mass dimensions - that is, only a few first terms
in the Hamiltonian can contribute to the energy at $O(e^2\lambda^2)$.
The same argument with a slight modification applies also to the sum with
one light particle in the intermediate state. Summing up the Coulomb
wave functions, one ends up with the Schwinger's Green function, with possible
insertions of higher-dimensional operators, folded by the light particle loop.
Further, one expands the Schwinger's function in powers of the external
field - then, each integral has the form already considered above, in the
context of the scattering problem. Using
the multipole expansion restores the power counting - the leading term
is determined by the naive power-counting rules, and the sub-leading
terms are explicitly given. Consequently, for this term in the perturbation
expansion one can also apply
the same power-counting arguments that were used in the terms with no
light field. Therefore, one has to examine only
few contributions from the lowest-order operators in ${\bf H}_{\rm I}$,
in order to obtain the energy-level shift in a given accuracy. In particular,
the coefficient $g_1$ is needed in the accuracy
$O(e^2)$, whereas it suffices to determine $g_2$
at tree level, and all higher-order terms can be ignored. To conclude,
in order to evaluate the energy-level shift at $O(e^2\lambda^2)$, it suffices
to work with the Hamiltonian given by Eqs.~(\ref{eq462})-(\ref{eq462_1}).
For a more detailed discussion of the power counting in the
bound-state calculations within the framework of non-relativistic effective
theories, we refer the interested reader to Refs.~\cite{QED,Soto}.

Let us now pass to the calculation of the ground-state energy-level shift.
The matrix element $\langle 1|{\bf H}_{\rm I}|1\rangle$ receives contribution
only from the second term in the interaction Hamilton operator,
\eq\label{eq465}
\langle 1|{\bf H}_{\rm I}|1\rangle=
-\frac{5e^2}{192\pi^2 M^3}\,\int d\nu({\bf p})\,
d\nu({\bf k})\,\psi_1^*({\bf p})\,\frac{4\pi\lambda}
{|{\bf p}-{\bf k}|^2}\,\psi_1({\bf k})
=-\,\frac{5e^2\lambda^2}{384\pi^2 M^3}\, .
\en

The second and fourth terms in the perturbative expansion~(\ref{eq463}) do
not contribute to the bound-state energy at $O(e^2\lambda^2)$.
In the third term, the summation over the Coulomb wave functions can be
carried out explicitly, resulting in Schwinger's Green
function~\cite{Schwinger}. Then, the contribution from this term to the
energy-level shift can be written as follows
\eq\label{eq466}
\frac{e^2}{4M^2}\int d\nu_d({\bf p})\,
d\nu_d({\bf k})\,{\psi_1^d}({\bf p})^*\psi_1^d({\bf k})
\int \, \frac{d\nu_d({\bf q})}{2|{\bf q}|}\,
(2\pi)^d\, G_S(E_1-M-|{\bf q}|;{\bf p}-{\bf q},{\bf k}-{\bf q})\, .
\en

An important remark is in order. In Eq.~(\ref{eq466}), one has to use
the Schwinger's function in $d$ dimensions, defined through the spectral sum
over $d$-dimensional wave functions (see Appendix~\ref{app:multi-C}).
Further, this function can be written as a sum of zero-Coulomb,
one-Coulomb and multi-Coulomb parts. In the first two, the generalization to
$d$ dimensions does not cause any problems, while the multi-Coulomb part
is ultraviolet convergent and can be considered at $d=3$.

The zero-Coulomb contribution to the bound state energy is given by
\eq\label{eq467}
\Delta E_{II,0}=\frac{e^2}{4M^2}\,
\int d\nu_d({\bf p})\, |\psi_1^d({\bf p})|^2
\int d\nu_d({\bf q})\,\,
\frac{1}{2|{\bf q}|}\,\frac{1}{E_1^d-M-|{\bf q}|
-\frac{({\bf p}-{\bf q})^2}{2M}}\, .
\en
In order to be consistent with matching in the scattering sector, one has to
carry out the multipole
expansion in this integral as well. The result at $O(e^2\lambda^2)$ then is
\eq\label{eq468}
\Delta E_{II,0}=-\frac{e^2}{2M^2}\, \bar L_M \int d\nu_d({\bf p})\,
|\psi_1^d({\bf p})|^2\,\frac{{\bf p}^2+2M(M-E_1^d)}{2M}\,
-\frac{e^2\lambda^2}{64\pi^2M^3}\,
\biggl(\ln\frac{\lambda^2}{M^2}-\frac{3}{2}\biggr)\, .
\en

The one-Coulomb contribution to the bound-state energy is calculated
analogously,
\eq\label{eq469}
\Delta E_{II,1}&=&-\frac{\pi e^2\lambda}{2M^3}
\int\, d\nu_d({\bf p})\, d\nu_d({\bf k})\,
\psi_1^d({\bf p})^*\psi_1^d({\bf k})\, \frac{1}{|{\bf p}-{\bf k}|^2}
\times
\\[2mm]
&\times&
\int\, d\nu_d({\bf q})\,\, \frac{1}{2|{\bf q}|}\,\,
\frac{1}
{\bigl(E_1^d-M-|{\bf q}|-\frac{({\bf p}-{\bf q})^2}{2M}\bigr)
\bigl(E_1^d-M-|{\bf q}|-\frac{({\bf k}-{\bf q})^2}{2M}\bigr)}\, .
\nonumber
\en
After carrying out the multipole expansion, we obtain
\eq\label{eq470}
&&\Delta E_{II,1}=\frac{e^2}{2M^2}\, \bar L_M
\int\, d\nu_d({\bf p})\,
|\psi_1^d({\bf p})|^2\,\frac{{\bf p}^2+2M(M-E_1^d)}{2M}\,
+\frac{e^2\lambda^2}{64\pi^2M^3}\,
\biggl(\, \ln\frac{\lambda^2}{M^2}-6+8\ln 2\, \biggr)\, .
\nonumber\\
&&
\en
Note that the divergent terms as well as terms non-analytic in $\lambda$,
cancel in the sum $\Delta E_{II,0}+\Delta E_{II,1}$.

In the multi-Coulomb part of the energy shift, as it was mentioned above,
one can directly perform the
limit $d\rightarrow 3$, since this contribution is ultraviolet and infrared
finite.
After re-scaling the integration variables, this piece is written as follows
\eq\label{eq471}
&&\Delta E_{II,m}=-\frac{8\pi^2e^2\lambda^2}{M^3}\int\,
d\nu({\bf u})d\nu({\bf v})
\frac{1}{(1+{\bf u}^2)^2}\frac{1}{(1+{\bf v}^2)^2}
\times
\\[2mm]
&&\times
\int\, d\nu({\bf w})
\frac{1}{|{\bf w}|\sqrt{1+|{\bf w}|}}\,\,
\int_0^1\frac{d\rho\rho^{-(1+|{\bf w}|)^{-1/2}}}
{F(\rho;|{\bf w}|;{\bf u},{\bf v})}\,\,
\frac{1}{1+|{\bf w}|+({\bf u}-\frac{\gamma}{2M}{\bf w})^2}\,\,
\frac{1}{1+|{\bf w}|+({\bf v}-\frac{\gamma}{2M}{\bf w})^2}\, ,
\nonumber
\en
where
\eq\label{eq472}
&&F(\rho;|{\bf w}|;{\bf u},{\bf v})=
({\bf u}-{\bf v})^2\rho+\frac{(1-\rho)^2}{4(1+|{\bf w}|)}
(1+|{\bf w}|+({\bf u}-\frac{\gamma}{2M}{\bf w})^2)
(1+|{\bf w}|+({\bf v}-\frac{\gamma}{2M}{\bf w})^2)\, ,
\nonumber\\
&&
\en
and $\gamma=\lambda/2$.
Taking the limit $\gamma\rightarrow 0$, one identifies this
 expression with the analogous contribution in the relativistic case,
Eqs.~(\ref{m-C-J}) and (\ref{J}). Finally, collecting all contributions to
the radiative energy-level shift, we obtain
\eq\label{eq473}
\Delta E_1=\langle 1|{\bf H}_{\rm I}|1\rangle+\Delta E_{II,0}+\Delta E_{II,1}+
\Delta E_{II,m}
=-\, \frac{e^2\lambda^2}{64\pi^2M^3}\biggl(\frac{16}{3}-8\ln 2 +J\biggr)
+\cdots\, ,
\en
which is exactly the result~(\ref{eq437}) obtained by using the relativistic
framework.

\setcounter{equation}{0}

\section{Summary and conclusions}
\label{sec:summary}

\begin{itemize}
\item[i)]
We have  considered  in this article the energy-levels of a massive
 scalar particle, which
moves in an external Coulomb field and interacts in addition with a
massless scalar particle  through a Yukawa interaction. This  has
allowed us to study in detail the non-relativistic effective Lagrangian
treatment\cite{Lepage} of the bound-state problem in Quantum Field Theory.
\item[ii)]
 We first considered the relativistic calculation of the energy
levels. The calculation
serves as a reference point for the non-relativistic calculation
performed later on - with the  exception of the non-renormalizable
interaction considered at the end of section II. Indeed, that
interaction  mimics the situation encountered in ChPT.
We have shown that - at lowest nontrivial
order in the relevant coupling constant $g$ -  bound-state observables
are ultraviolet finite, once the Green functions in the scattering
sector are made finite by an appropriate choice of the counterterms.
Although we have not proven it, we have no doubt that this will remain
true to all orders on $g$.
\item[iii)]
A systematic framework for the perturbative calculation of bound-state
observables was then developed in the non-relativistic sector,
based on Feshbach's \cite{Feshbach} technique. At the end,
one recovers the conventional Rayleigh-Schr\"{o}dinger
perturbative expansion of the energy-levels. The method allows one in
addition to calculate the position of the pole on the second Riemann
sheet, in case that the bound state turns out to be unstable.
\item[iv)]
Although the external field problem may sound trivial at first, actual
calculations in the non-relativistic theory reveal  that this is not the case.
Indeed, applying the Rayleigh-Schr\"odinger perturbative expansion,
one has to verify that the ultraviolet divergences in the matrix elements
of high-dimensional operators cancel, provided that  all contributions
at a given order in coupling constant are summed up.
Stated  differently, the non-relativistic expansion does not
generate additional ultraviolet divergences, despite the fact that
the non-relativistic Lagrangian contains  terms with arbitrarily
high powers of spatial derivatives.
\item[v)]
We have explicitly checked in an example that the off-shell ambiguity in
the effective theory - that can be traced to the presence of  terms in
the Lagrangian which vanish upon use of the equation of motion or
can be removed by field redefinitions - does not affect the energy-level.
[Similar results were recently obtained in the context of NRQCD~\cite{EOM1}
and in the non-relativistic effective field theory approach to
 the deuteron~\cite{EOM2}.]
\item[vi)]
Next, we considered  radiative corrections  in the non-relativistic
sector in some detail. The construction of the relevant effective
Lagrangian is based on the following idea. First, one constructs a
Lagrangian that agrees at tree level with the relativistic amplitude
 to all orders in the low-energy  expansion.
This guarantees that the correct cut structure is obtained at one-loop
level, in the sense that the absorptive parts of the relativistic
theory  are reproduced. Loops then simply amount to close two of the external
lines in the tree graphs - therefore, we expect that  the
non-relativistic $S$-matrix
elements do have the correct non-analytic structure at one-loop order.
\item[vii)]
We have investigated the
two-point function of the heavy particle at second order in the Yukawa
coupling. It was demonstrated that - at zero three-momentum - the
multipole expansion of the Feynman
integral reproduces the non-analytic parts of the original expression, order
by order in the expansion. The polynomial part of the
integral - which leads to the breakdown of counting rules -
is modified by the multipole expansion.  This merely amounts to a change
 of the renormalization prescription and does not affect
 physical observables.
\item[viii)]
In addition, we have investigated the structure of the two-point
function in the threshold region, and
have shown how the correct pole structure emerges from the multipole
expansion. Eq. (\ref{eq527}) displays the expected relation between the
non-relativistic and the
relativistic self-energies at one-loop order, but at any order in the
$v$-expansion. That relation presumes that one has properly chosen
the counterterms.
\item[ix)]
We have considered the renormalization procedure
for the non-relativistic effective
theory, including the subtleties that arise due to the
presence of zero-mass particles. In particular, the renormalization of two
and three-point functions in the non-relativistic theory was treated in
detail. Further, we have  demonstrated that - at leading order in the
low-energy expansion - a procedure which
utilizes  no-scale arguments in the dimensional regularization is
equivalent to the standard  renormalization scheme.
\item[x)]
Constructing finally the non-relativistic Hamiltonian in the presence
of the light field, we have  verified that it reproduces -
at first nontrivial order in the Yukawa coupling - the relativistic
expression for the energy-level of the ground state.
\end{itemize}

In conclusion, we find that dimensional regularization, together with
the multipole expansion~\cite{QED,Manohar,Beneke,Griesshammer} and Feshbach's
representation
\cite{Feshbach}  of the Green functions,
allows  a sound evaluation of energy-levels in the non-relativistic models
considered. Furthermore, with the same technique, one can determine
the position of the poles on the second Riemann sheet for decaying
states \cite{Bern1,Bern2}. No use was made of a representation
of the Lagrangian in terms
of expanded fields \cite{Manohar,Griesshammer} in order to perform these
calculations.

\subsection*{Acknowledgments}

We are grateful to M.~Beneke, H.~Griesshammer,
M.A.~Ivanov, T.~Kinoshita, P.~Labelle, G.P.~Lepage,
H.~Leutwyler, V.E.~Lyubovitskij, K.~Melnikov, P.~Minkowski,
M.~Nio, H.~Sazdjian and J.~Soto for useful discussions.
This work was supported in part by the Swiss National Science
Foundation, and by TMR, BBW-Contract No. 97.0131  and  EC-Contract
No. ERBFMRX-CT980169 (EURODA$\Phi$NE).

\appendix

\renewcommand{\thesection}{\Alph{section}}
\renewcommand{\theequation}{\Alph{section}\arabic{equation}}

\section{One-Coulomb contribution to the energy-level shift}
\label{app:1-C}
\setcounter{equation}{0}

In order to calculate the one-Coulomb contribution to the energy shift,
we first combine the denominators in Eq.~(\ref{eq427}) by using Feynman
parameters. Then, one obtains
\eq\label{D1}
&&\Gamma(E_1;{\bf p},{\bf k})=\frac{1}{16\pi^2}\,\,\int_0^1\, dx\,\,
\frac{\ln g(x)-\ln f(x)}{g(x)-f(x)}\, ,
\nonumber\\[2mm]
&&g(x)=M^2+x(1-x)({\bf p}-{\bf k})^2\, ,\quad\quad
f(x)=x(\gamma^2+{\bf p}^2)+(1-x)(\gamma^2+{\bf k}^2)\, .
\en

Next, we substitute Eq.~(\ref{D1}) into Eq.~(\ref{eq426}), and re-scale the
integration variables according to ${\bf p}=\gamma{\bf u}$,
${\bf k}=\gamma{\bf v}$. We obtain
\eq\label{D2}
F_{II,1}=8z^2(\ln z^2-2\ln 2-1)\, I_1+8z^2\, I_2+o(z^2)\, ,
\en
where
\eq\label{D3}
I_1=\int d\nu({\bf u})d\nu({\bf v})\,\frac{1}{(1+{\bf u}^2)^2}\,
\frac{1}{(1+{\bf v}^2)^2}\, \frac{1}{|{\bf u}-{\bf v}|^2}=\frac{1}{256\pi^2}
\, ,
\en
and
\eq\label{D4}
&&I_2=\int d\nu({\bf u})d\nu({\bf v})\,\frac{1}{(1+{\bf u}^2)^2}\,
\frac{1}{(1+{\bf v}^2)^2}\, \frac{1}{|{\bf u}-{\bf v}|^2}\,
\frac{(1+{\bf u}^2)\ln(1+{\bf u}^2)-(1+{\bf v}^2)\ln(1+{\bf v}^2)}
{{\bf u}^2-{\bf v}^2}\, .
\nonumber\\
&&
\en
In order to calculate $I_2$, it is convenient to use the integral
representation
\eq\label{D5}
&&\frac{(1+{\bf u}^2)\ln(1+{\bf u}^2)-(1+{\bf v}^2)\ln(1+{\bf v}^2)}
{{\bf u}^2-{\bf v}^2}=
\int_0^\infty dx\biggl(\frac{1}{x+1}-\frac{x}{(x+1+{\bf u}^2)(x+1+{\bf v}^2)}
\biggr)\, ,
\nonumber\\
&&
\en
and to introduce the Fourier transform of $|{\bf u}-{\bf v}|^{-2}$,
\eq\label{D6}
\frac{1}{|{\bf u}-{\bf v}|^2}=\frac{1}{4\pi}\,\int d^3{\bf r}\,
\frac{e^{i({\bf u}-{\bf v}){\bf r}}}{r}\, .
\en
After this, all integrations in $I_2$ can be performed analytically, resulting
in
\eq\label{D7}
I_2=-\frac{5}{128\pi^2}\,\biggl(\frac{1}{2}-\ln 2\biggr)\, .
\en

Substituting Eqs.~(\ref{D3}) and (\ref{D7}) into Eq.~(\ref{D2}),
we arrive at the final result displayed in Eq.~(\ref{eq430}).

\section{Multi-Coulomb contributions to the energy-level shift}
\label{app:multi-C}
\setcounter{equation}{0}

In this Appendix, we present the calculation of the multi-Coulomb contribution
to the energy-level shift. In this calculation, a straightforward use of
Schwinger's representation encounters a
difficulty, since in the integration region $-\infty<q^0<\infty$ the variable
$\nu$ in Eq.~(\ref{eq:2-14}) can become larger than $1$. In order to
overcome this difficulty, we use the spectral representation for the
Schwinger's function in terms of the solutions $|n{\rangle}_d$ of the
non-relativistic Schr\"{o}dinger equation with the Coulomb
potential~(\ref{eq221}).
By using this representation, the integration over the variable $q^0$ can be
 performed with the use of Cauchy's theorem. The whole
contribution from the diagrams of the type II to the self-energy part in
$D$ dimensions can be written in the following form
\eq\label{eq431}
&&\Sigma^{II}(E_1^d;{\bf p},{\bf k})=-ie^2\sum_n\int\frac{d^Dq}{(2\pi)^D}
\frac{1}{q^2+i0}\,\,
\frac{\langle{\bf p}-{\bf q}|n{\rangle_d}~
{_d\langle} n|{\bf k}-{\bf q}\rangle}
{(E_1^d-q^0)^2-(E_n^d)^2+i0}
\nonumber\\[2mm]
&=&e^2\sum_n\int d\nu_d({\bf q})\,\,
\frac{\langle{\bf p}-{\bf q}|n{\rangle_d}~
{_d\langle} n|{\bf k}-{\bf q}\rangle}
{4E_n^d|{\bf q}|}\,\,
\biggl(\frac{1}{E_n^d-E_1^d+|{\bf q}|}+\frac{1}{E_n^d+E_1^d+|{\bf q}|}\biggr)
\, ,
\en
where the sum runs both over the discrete and the continuous
spectra of the unperturbed Coulomb Hamilton operator.

The representation (\ref{eq431}) now includes all contributions
from the diagrams of the type II, apart from the mass counterterm
$-\delta M^2$. To single out from this expression the low-energy
contribution which contributes at the leading order in $\lambda$, we
introduce an additional integration over the auxiliary variable $x$,
using the following identities
\eq\label{eq432}
\frac{1}{E_n^d(E_n^d-E_1^d+|{\bf q}|)}&=&\frac{2\theta(E_1^d-|{\bf q}|)}
{(E_n^d)^2-(E_1^d-|{\bf q}|)^2}
-\frac{2}{\pi}\int_0^\infty\,
\frac{dx\,(E_1^d-|{\bf q}|)}{(x^2+(E_1^d-|{\bf q}|)^2)(x^2+(E_n^d)^2)}\, ,
\nonumber\\[2mm]
\frac{1}{E_n^d(E_n^d+E_1^d+|{\bf q}|)}&=&\frac{2}{\pi}\int_0^\infty\,
\frac{dx\,(E_1^d+|{\bf q}|)}{(x^2+(E_1^d+|{\bf q}|)^2)(x^2+(E_n^d)^2)}\, .
\en

If now one substitutes Eq.~(\ref{eq432}) into Eq.~(\ref{eq431}),
 one can  single out the piece corresponding the multi-Coulomb exchanges.
[In this piece, one can safely put $D=4$, $E_1^d=E_1$, etc,
since there are no ultraviolet divergences.]
As far as this contribution is concerned,
it is seen that only the part of this expression which does not involve
the additional integration over $x$ (the one coming from the first
term in the r.h.s. of the first line of Eq.~(\ref{eq432})), contributes
at order $O(\lambda^2)$. Other terms can be  proven to contribute at
order $O(\lambda^4)$. The reason for the use of the above trick is the
following. One can  observe from Eq.~(\ref{eq432}) that now the
variables which serve as an ``energy variable'' in the Schwinger's
representation: $\tilde\epsilon=((E_1-|{\bf q}|)^2-M^2)/(2M)$ for the
term without $x$ integration, and $\tilde\epsilon(x)=-(x^2+M^2)/(2M)$
for the terms with the $x$ integration, vary in the regions
$-M/2\leq\tilde\epsilon\leq -\lambda^2/(8M)$ and
$-\infty<\tilde\epsilon(x)\leq -M/2$, respectively. The corresponding
values of the parameter $\nu$ in the Schwinger's representation never
exceed $1$ and, consequently, the integral over $d\rho$ in Eq.~(\ref{eq:2-14})
always converges.

According to the above discussion, with the use of the Schwinger's
representation one can now write down the part of the multi-Coulomb
contribution to the heavy particle self-energy graph, which gives the
leading-order contribution to the energy shift
\eq\label{eq433}
&&\Sigma^{II,m}(E_1;{\bf p}-{\bf q},{\bf k}-{\bf q})=e^2
\int^{|{\bf q}|\leq E_1}d\nu({\bf q})\,
\frac{1}{2|{\bf q}|}\,\,
\frac{2\pi\lambda^2 I(E_1-|{\bf q}|;{\bf p}-{\bf q},{\bf k}-{\bf q})}
{(M^2-(E_1-|{\bf q}|)^2)^{1/2}}\times
\nonumber\\[2mm]
&&\hspace*{1.cm}\times
\frac{1}{M^2+({\bf p}-{\bf q})^2-(E_1-|{\bf q}|)^2}\,\,
\frac{1}{M^2+({\bf k}-{\bf q})^2-(E_1-|{\bf q}|)^2}+\cdots\, ,
\en
where the quantity $I$ is given by Eq.~(\ref{eq:2-14}),
and the ellipsis stands for  terms which generate higher-order contributions
in $\lambda$.

Substituting Eq.~(\ref{eq433}) into the expression
of the energy-level shift and re-scaling the integration variables
${\bf p}=\gamma{\bf u}$, ${\bf k}=\gamma{\bf v}$,
${\bf q}=\gamma^2{\bf w}/(2M)$,
one arrives at the Eqs.~(\ref{m-C-J}) and (\ref{J}).

\section{Use of the equation of motion}
\label{app:EOM}
\setcounter{equation}{0}
\subsection{The effective Lagrangian (\ref{eq:2-23})}
The original form of the Lagrangian
is obtained by expanding the
propagator $D_-^{-1}$ and  the differential operator $d$ - that
appear in the operator ${\cal D}$ in (\ref{eq:2-19}) - in inverse powers
of the heavy mass,
\eq
  \label{A1}
  D_-^{-1}&=&-\left\{d_t+M+\sqrt{M^2-\Delta}\right\}^{-1}\!\!\!
  =-\frac{1}{2M}\left\{1-\frac{d_t}{2M} +
    \frac{1}{4M^2}(d_t^{\, 2}+\Delta)+\ldots\right\}\, ,
\nonumber\\[2mm]
  d&=& \frac{1}{\sqrt{2M}}\left\{1+\frac{\Delta}{4M^2}+\ldots\right\}
  \sem \quad\quad d_t\doteq i\partial_t-M\per
\en
The result is
\eq
  \label{A2}
  {\cal L}_{NR}&=&h^\dagger{\cal D}h=
  h^\dagger D_+h + \sum_{n=1}^\infty \frac{1}{(2M)^n}{\cal L}^{(n)}\scs
  \nonumber\\[2mm]
  {\cal L}^{(1)}&=& \lambda h^\dagger\phi h \scs
  {\cal L}^{(2)} = 0 \scs
\nonumber\\[2mm]
  {\cal L}^{(3)}&=&  \lambda h^\dagger(\phi\triangle + \triangle\phi)h +
\lambda^2h^\dagger\phi^2 h\scs
  {\cal L}^{(4)} = -\lambda^2 h^\dagger\phi d_t\phi h\co
  \nonumber \\[2mm]
  {\cal L}^{(5)}&=&  \lambda h^\dagger\left(\triangle\phi\triangle +
    \frac{5}{2}\phi\triangle^2 +
    \frac{5}{2}\triangle^2\phi\right) h\!
\nonumber\\[2mm]
                & +& \!{\lambda}^2 h^\dagger\left(\phi^2\triangle \!+\!
    \triangle\phi^2 + 3\phi\triangle\phi  + \phi d_t^{\, 2}\phi \right) h+
  {\lambda}^3h^\dagger\phi^3 h.
\en
The time derivatives of second and higher order may be eliminated in the
following manner. First, we write the Lagrangian as
\eq
  \label{A3}
  {\cal{L}}_{NR}=h^\dagger Dh\co\quad\quad
  D=d_t+A_0+A_1d_t+\sum_{n \geq 2} A_n d_t^n\per
\en
The operator
$A_0$ contains Laplacians and external fields $\phi$, and
$A_{1,2,3\ldots}$ contain at least two external fields. The
two-point function is obtained by evaluating the classical action
\eq
  \label{A4}
  S=\int d^4x ({\cal L}_{NR} +h^\dagger j +j^\dagger h)
\en
at the solution to the classical equation of motion,
\eq
  \label{A5}
  Dh=-j\co
\en
as a result of which
\eq
  \label{A6}
  S=\int d^4x j^\dagger h\co h=-D^{-1}j\per
\en
We eliminate the time derivatives successively from
Eq.~(\ref{A5}). First, we consider the quantity
\eq
\bar{S}=\int d^4x j^\dagger \bar{h}\co
\en
where
\eq
\bar{h}=-(\frac{1}{1+A_1}D)^{-1}j = -D^{-1}j -D^{-1}A_1j\per
\en
The second term does not contribute to on-shell matrix elements, because on
one of the external legs, there is no pole: the source $j$ couples directly to
the external fields $\phi$ present in $A_1$. We conclude that
the quantity $\bar{S}$ contains the same on-shell matrix elements as
the original classical action $S$.  Stated otherwise, one may use,
instead of (\ref{A3}), the Lagrangian
\eq
  \label{A10}
  {\cal L}_{NR}=h^\dagger D_1h\co
\en
with
\eq
  \label{A8}
  D_1&=& \frac{1}{1+A_1}D\nonumber\\
      &=&d_t+B_0+\sum_{m\geq 2}B_m d_t^m\sem
  B_m=\frac{1}{1+A_1}A_m\scs m=0,2,3,\ldots\, .
\en
The on-shell matrix elements are obtained by inserting in (\ref{A6})
the solution of
\eq\label{A11}
D_1h=-j\per
\en
By acting with $B_2d_t$ on (\ref{A11}), solving for $B_2d_t^2$ and
proceeding as before, one may eliminate $B_2d_t^2$,
and similarly for all higher time derivatives.

\subsection{The effective Lagrangian (\ref{Lagrangian-ell})}
The main observation is the fact that,
 as far as tree graphs in the one heavy particle sector are
concerned, we may  perform in the classical action (\ref{A4})
the replacement $\lambda \phi\rightarrow eD_0^{-1}f$, where $f$ is the
external field associated with the light particle, and $\Box
 D_0^{-1}(x)=\delta^4(x)$. Therefore, the problem of eliminating
higher order time
 derivatives is reduced to the one  the previous discussion, except
 that now, the external field is time-dependent. The
 time derivatives act  on the heavy and on the
  external field. The ones acting on the heavy field may be removed as
 is described above. As for the time derivatives
 on  the  external field, one  uses $\Box D_0^{-1}(x)=\delta^4(x)$ and can thus
 eliminate all of them, except those of first order.

\subsection{Re-scaling the heavy field}
Constructing the Hamiltonian, it is useful to rely on  a Lagrangian whose
kinetic term is properly normalized. This is not the case for the one
displayed in
(\ref{eq61}) due to the presence of the counterterms. Since these
 are needed only at tree level,
one may again use the above described manipulations to eliminate them.
 However, some of the unwanted time derivatives
differ from the ones  in (\ref{A3}), to the extent that
the operators $A_n$ need now not contain external fields  - an
example is the counterterm proportional to $f_1$ in (\ref{eq61}). As a
result of this,
the manipulations described after Eq.~(\ref{A3}) lead to Green
functions that differ from the original ones also on the mass-shell
- to quote an example,  the residue of the two-point function is
be modified.
 However, one can  convince oneself that, in the
calculation of the $S$-matrix elements,  this change in the two-point
function is exactly compensated by a corresponding change of the
coupling constants in
the effective Lagrangian. As a result of this, the manipulations
leading from (\ref{A3}) to (\ref{A8}) generate  a Lagrangian that
leaves the $S$-matrix elements untouched.

\section{Connection of the resolvent with the two-point function}
\setcounter{equation}{0}
\label{app:resolvent}
In this Appendix we derive Eq.~(\ref{eq:resolvent-2pf}) of
section~\ref{sec:NRex}, which gives the connection of the
resolvent
\eq
  \langle{\bf p}|\frac{1}{E-{\bf H}}|{\bf k}\rangle
\en
with the two-point function
\eq
  G_{NR}(E;{\bf p}, {\bf k}) = i\int dt d^3{\bf x} d^3{\bf y}
 e^{iEt- i{\bf p}{\bf x} +
  i {\bf k} {\bf y}}\langle 0|T h(t,{\bf x})h^\dagger(0,{\bf y})|0\rangle .
\en

Expressing the creation and annihilation operators of one-particle states in
terms of the fields, Eq.~(\ref{aadagger}), and using the representation
\eq
  \frac{1}{E-{\bf H}} = \frac{1}{i}\int_0^\infty dt e^{i(E-{\bf H})t},
\en
which is valid for ${\rm Im}~E>0$, as well as the fact that ${\bf H}$ generates
translations in time
\eq
  h(t,{\bf x}) = e^{i{\bf H} t}h(0,{\bf x}) e^{-i{\bf H} t},
\en
we get
\eq
  \langle{\bf p}|\frac{1}{E-{\bf H}}|{\bf k}\rangle =
-\int_0^\infty dt \int d^3{\bf x} d^3{\bf y}
e^{iEt-i{\bf p}{\bf x} +i{\bf k}{\bf y}}
i\langle 0|e^{-i{\bf H} t}h(t,{\bf x}) h^\dagger(0,{\bf y})|0\rangle .
\en
We assume that the full Hamilton operator annihilates the vacuum ${\bf
H}|0\rangle=0$.
Since $t>0$, we may replace the expectation value of the ordinary product with
that of the time-ordered product of fields without changing anything. Finally,
observing
that $\langle 0|h^\dagger(0,{\bf y}) = 0$, we can extend the $t$-integration to
$-\infty$ and find
\eq
  \langle{\bf p}|\frac{1}{E-{\bf H}}|{\bf k}\rangle = -G_{NR}(E;{\bf p},{\bf
k}).
\en
Note that the last step is a consequence of the non-relativistic nature of the
theory: the number of particles associated with the field $h$ cannot change.

\newpage

\begin{center}
{\bf FIGURE CAPTIONS}
\end{center}

\begin{figure}[h]
\epsfxsize=9cm
\epsfysize=4cm
\vspace*{.4cm}
\begin{picture}(50,15) \end{picture}
\epsffile{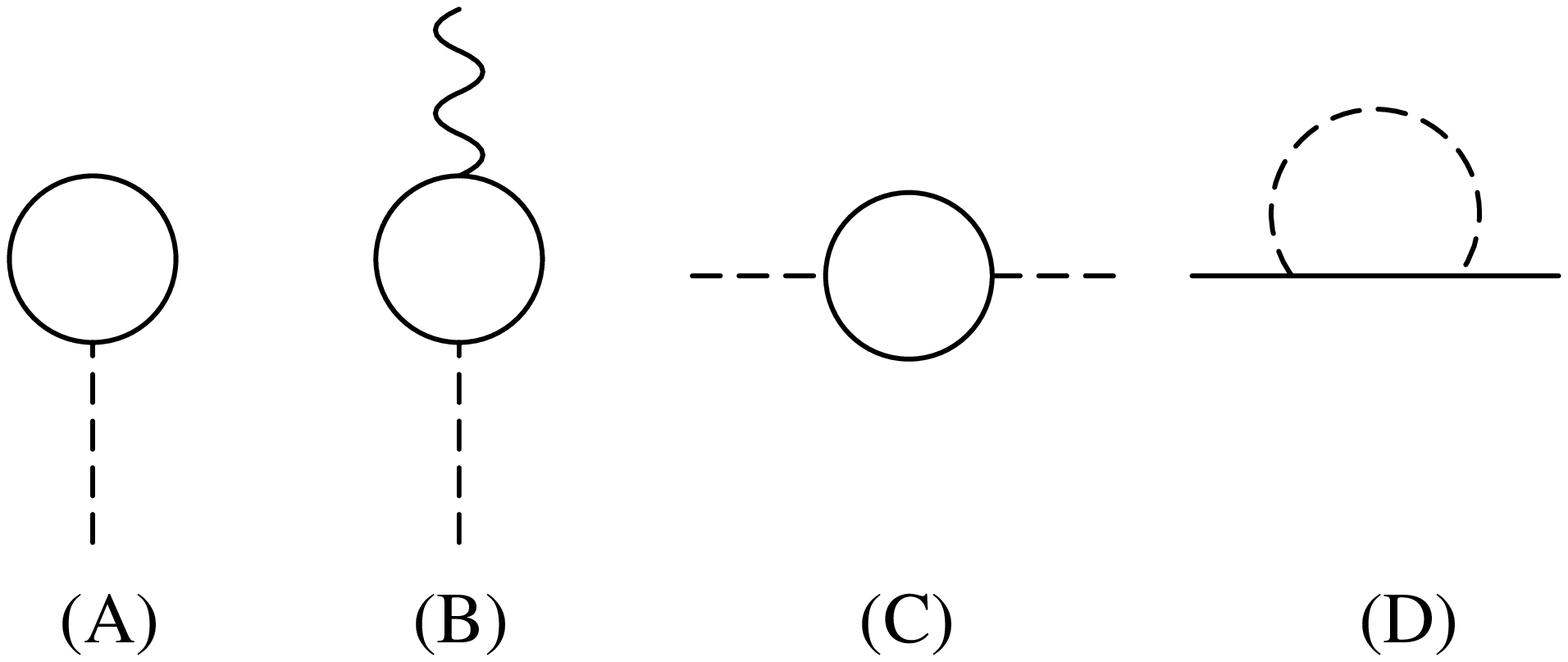}\vspace*{.4cm}
\caption{}\label{UV}
\end{figure}

\noindent {\bf FIG.~\ref{UV}.}
The ultraviolet divergent diagrams in the relativistic theory at one loop.
The heavy (light) particle is denoted by a solid (dashed) line,
 and the wavy line represents  the external field.
(A) Tadpole diagram,
(B) the interaction of the light field with the external field
through a  heavy  loop,
(C) the light particle self-energy graph,
(D) the heavy particle self-energy graph.

\vspace*{.3cm}

\begin{figure}[h]
\epsfxsize=8cm
\epsfysize=3cm
\begin{picture}(50,15) \end{picture}
\epsffile{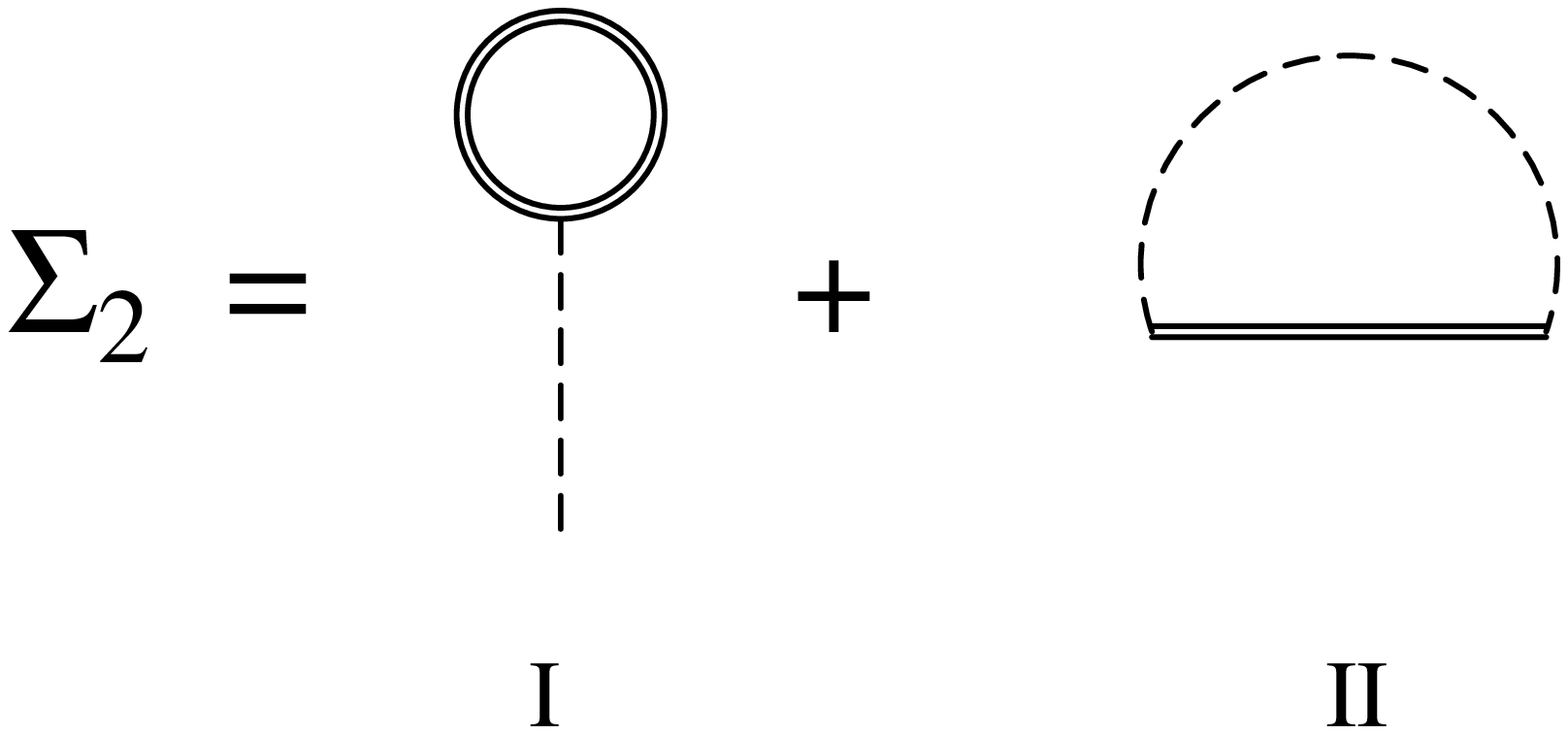}\vspace*{.4cm}
\caption{}\label{I-II}
\end{figure}

\noindent {\bf FIG.~\ref{I-II}.}
The self-energy diagrams of the heavy  particle moving in an
external field (relativistic theory, order $e^2$). The counterterm
contributions are not not shown. The double line corresponds to the
propagator $\bar G(x,y)$, and the dashed line denotes a light particle.

\vspace*{.3cm}

\begin{figure}[h]
\epsfxsize=8cm
\epsfysize=4cm
\begin{picture}(50,15) \end{picture}
\epsffile{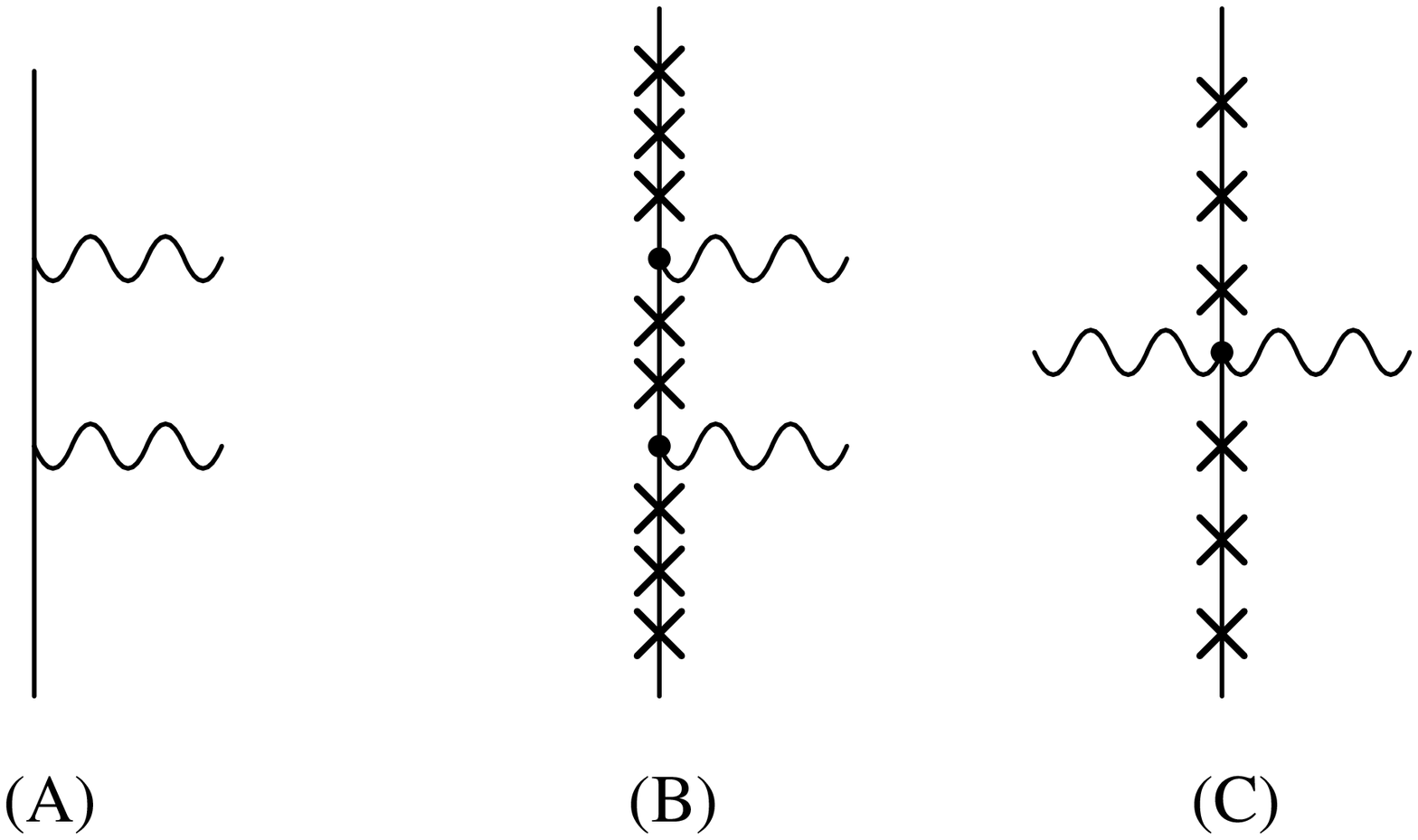}\vspace*{.4cm}
\caption{}\label{scat-ex}
\end{figure}

\noindent {\bf FIG.~\ref{scat-ex}.}
Scattering of the heavy particle in the external field
at order $\lambda^2$. The heavy  particle is denoted by
a solid line, and the wavy line represents  the external field.
Graph (A) contributes to  the Green function  in the relativistic
theory, whereas (B) and (C) contribute in the non-relativistic formulation.
Crosses denote mass insertions.

\vspace*{.3cm}

\begin{figure}[h]
\epsfxsize=6cm
\epsfysize=6cm
\begin{picture}(50,15) \end{picture}
\hspace*{2.5cm}\epsffile{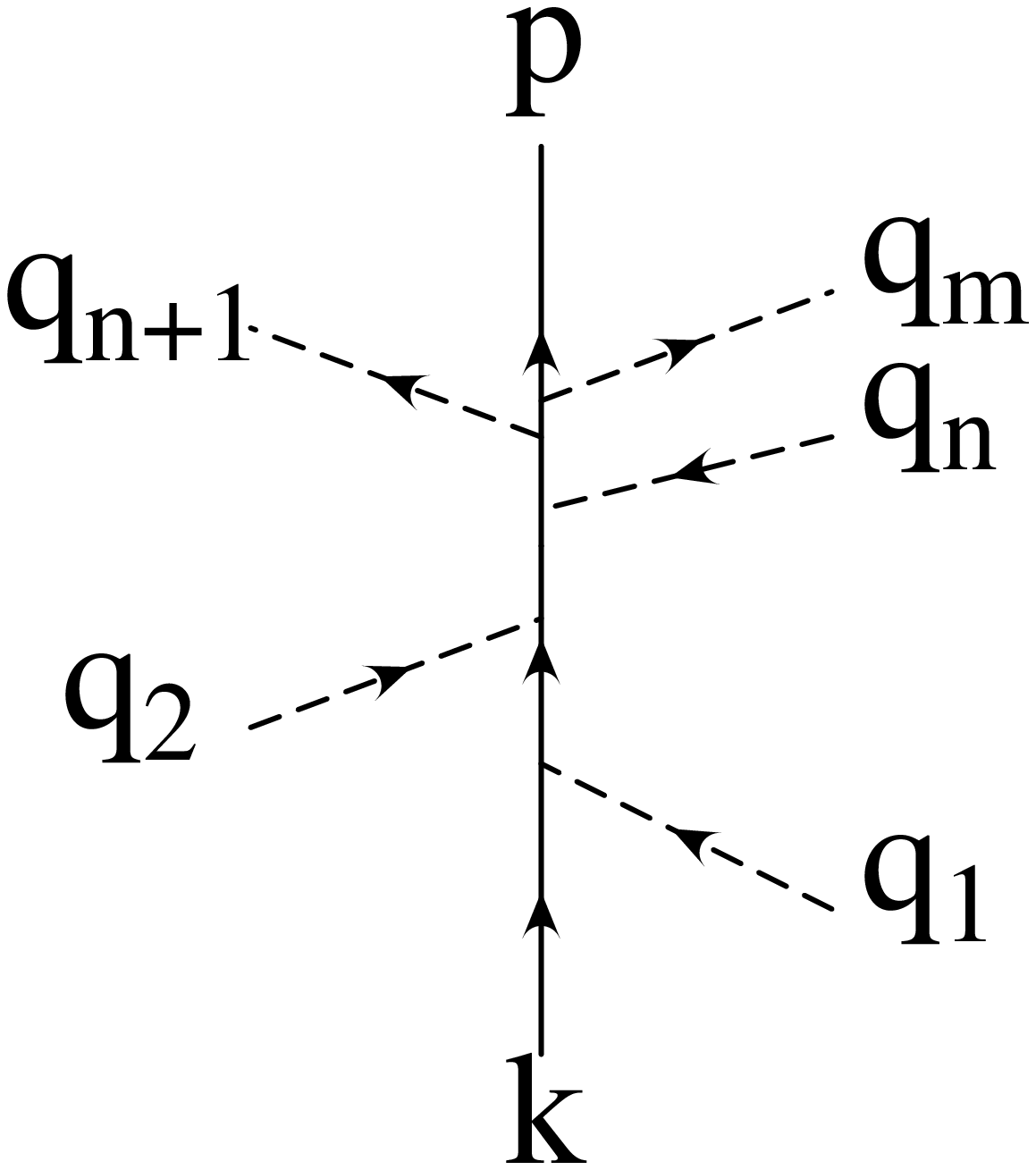}\vspace*{.4cm}
\caption{}\label{scat-ell}
\end{figure}

\noindent {\bf FIG.~\ref{scat-ell}.}
The scattering process $k+q_1+\cdots+q_n\rightarrow p+q_{n+1}+\cdots+q_m$
 in the relativistic theory at tree level.
The heavy (light) particle is denoted by a solid (dashed) line.

\vspace*{.3cm}

\begin{figure}[h]
\epsfxsize=14cm
\epsfysize=2.5cm
\begin{picture}(40,15) \end{picture}
\epsffile{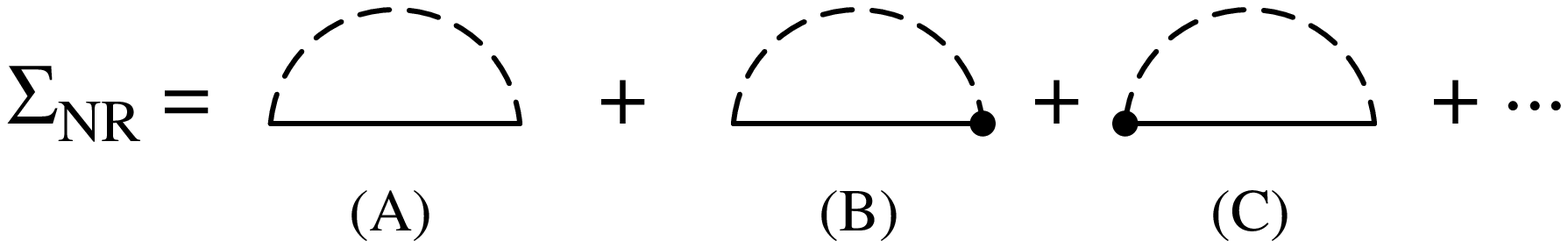}\vspace*{.4cm}
\caption{}\label{figselfnr}
\end{figure}

\noindent {\bf FIG.~\ref{figselfnr}.}
The self-energy graph of the heavy particle at order $e^2$,
non-relativistic theory.
The heavy (light) particle is denoted by a solid (dashed) line.
(A) The lowest-order graph with non-derivative vertices,
(B) and (C) are graphs with one derivative vertex, denoted by a filled circle.

\vspace*{.3cm}

\begin{figure}[h]
\epsfxsize=5cm
\epsfysize=2.5cm
\begin{picture}(40,15) \end{picture}
\hspace*{3.5cm}\epsffile{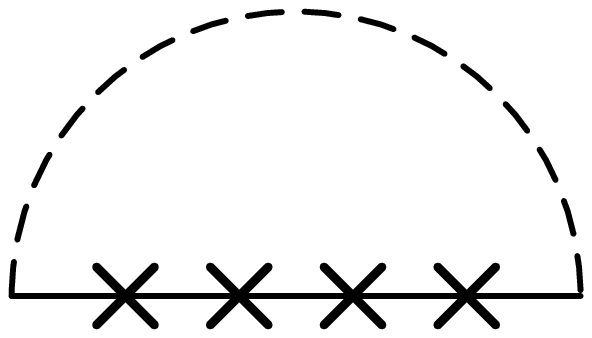}\vspace*{.4cm}
\caption{}\label{insertions}
\end{figure}

\noindent {\bf FIG.~\ref{insertions}.}
 Mass  insertions (indicated  by crosses) in the self-energy
 of the heavy particle, non-relativistic theory.
 The heavy (light) particle is denoted by a solid (dashed) line.

\vspace*{.3cm}

\begin{figure}[h]
\epsfxsize=12cm
\epsfysize=3cm
\begin{picture}(10,15) \end{picture}
\epsffile{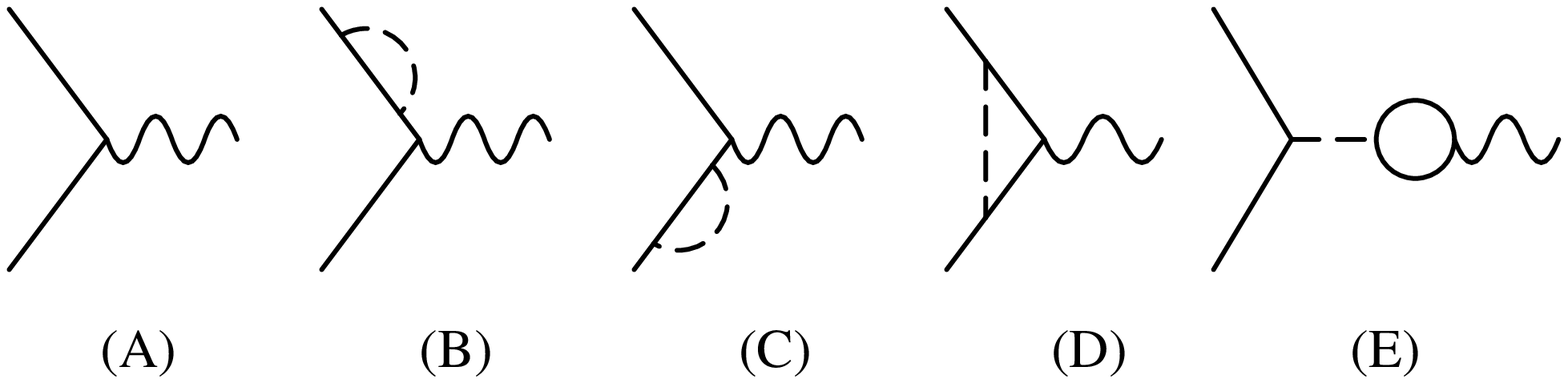}\vspace*{.4cm}
\caption{}\label{figvertex}
\end{figure}

\noindent {\bf FIG.~\ref{figvertex}.}
Scattering of the heavy particle in the external field at
order $e^2\lambda$, relativistic theory.
The heavy (light) particle is denoted by a solid (dashed) line,
 and the wavy line represents  the external field.
Diagrams containing counterterms are not shown.

\vspace*{.3cm}

\begin{figure}[h]
\epsfxsize=12cm
\epsfysize=3cm
\begin{picture}(10,15) \end{picture}
\epsffile{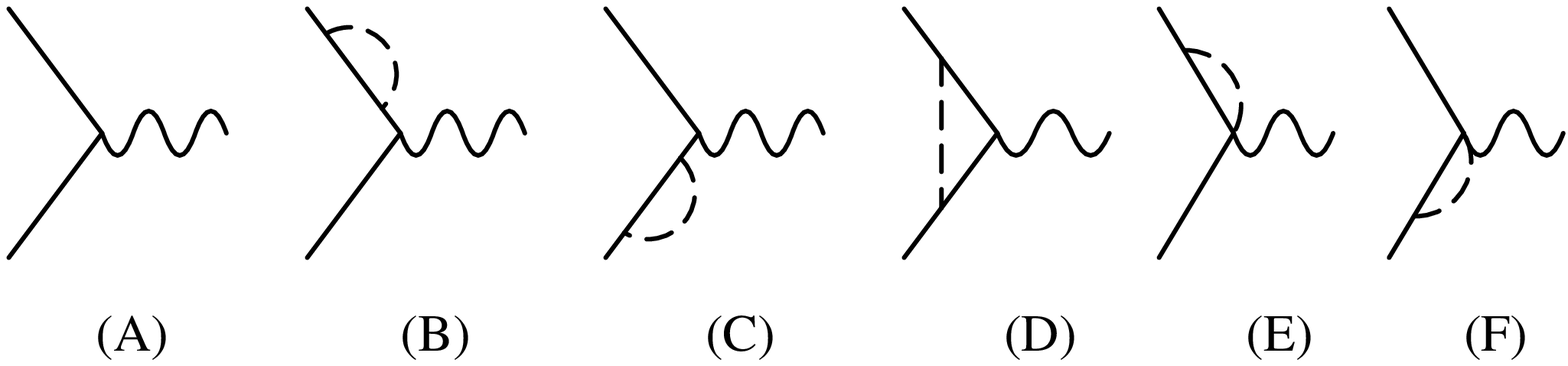}\vspace*{.4cm}
\caption{}\label{figvertex-NR}
\end{figure}

\noindent {\bf FIG.~\ref{figvertex-NR}.}
Scattering of the heavy particle in the external field at
order $e^2\lambda$, non-relativistic theory.
The heavy (light) particle is denoted by a solid (dashed) line,
 and the wavy line represents  the external field.
Diagrams containing counterterms are not shown.

\newpage

\vspace*{2.cm}

\vspace*{2.cm}

\newpage

\vspace*{2.cm}

\vspace*{2.cm}

\vspace*{2.cm}
\newpage

\vspace*{2.cm}

\clearpage


\begin{thebibliography}{99}

\bibitem{pipi-NR}
P.~Labelle and K.~Buckley, preprint hep-ph/9804201;
X.~Kong and F.~Ravndal, Phys. Rev. {\bf D~59}, 014031 (1999);
{\it ibid} {\bf D~61}, 077506 (2000);
B.R.~Holstein, Phys. Rev. {\bf D~60}, 114030 (1999);
D.~Eiras and J.~Soto, preprint hep-ph/9905543.

\bibitem{Bern1}
A.~Gall, J.~Gasser, V.E.~Lyubovitskij, and A.~Rusetsky
Phys. Lett. {\bf B~462}, 335 (1999).

\bibitem{Bern2}
J.~Gasser, V.E.~Lyubovitskij, and A.~Rusetsky,
Phys. Lett. {\bf B~471}, 244 (1999).


\bibitem{Lepage}
W.E.~Caswell and G.P.~Lepage, Phys. Lett. {\bf B~167}, 437 (1986).

\bibitem{pipi-rel}
H.~Jallouli and H.~Sazdjian, Phys. Rev. {\bf D~58}, 014011 (1998);
H.~Sazdjian, preprint hep-ph/9809425;
V.E.~Lyubovitskij and A.G.~Rusetsky, Phys. Lett. {\bf B~389}, 181 (1996);
V.E.~Lyubovitskij, E.Z.~Lipartia, and A.G.~Rusetsky,
JETP Lett. {\bf 66}, 783 (1997);
M.A.~Ivanov, V.E.~Lyubovitskij, E.Z.~Lipartia, and A.G.~Rusetsky,
Phys. Rev. {\bf D~58}, 094024 (1998).

\bibitem{QED}
P.~Labelle, PhD thesis, Cornell University (1994);
T.~Kinoshita and M.~Nio, Phys. Rev. {\bf D~53}, 4909 (1996);
Phys. Rev. {\bf D~55}, 7267 (1997);
P.~Labelle, S.M.~Zebarjad, and C.P.~Burgess,
Phys. Rev. {\bf D~56}, 8053 (1997);
P.~Labelle, Phys. Rev. {\bf D~58}, 093013 (1998).

\bibitem{Soto}
A.~Pineda and J.~Soto, Phys. Rev. {\bf D~59}, 016005 (1999).

\bibitem{Czarnecki}
K.~Melnikov and A.~Yelkhovsky, Phys. Lett. {\bf B 458}, 143 (1999);
A.~Czarnecki, K.~Melnikov, and A.~Yelkhovsky,
Phys. Rev. {\bf A 59}, 4316 (1999);
Phys. Rev. Lett. {\bf 82}, 311 (1999);
A.~Czarnecki, G.P.~Lepage, and W.J. Marciano,
Phys. Rev. {\bf D 61}, 073001 (2000).

\bibitem{NRQCD-HQET}
G.T.~Bodwin, E.~Braaten, and G.P.~Lepage, Phys. Rev. {\bf D~51}, 1125
(1995).

\bibitem{Manohar}
M.~Luke and A.V.~Manohar, Phys. Rev. {\bf D~55}, 4129 (1997);
A.V.~Manohar, Phys. Rev. {\bf D~56}, 230 (1997);
B.~Grinstein and I.Z.~Rothstein, Phys. Rev. {\bf D~57}, 78 (1998);
M.E.~Luke, A.V.~Manohar, and I.Z.~Rothstein, Phys. Rev. {\bf D~61}, 074025
(2000).

\bibitem{Beneke}
M.~Beneke and V.A.~Smirnov, Nucl. Phys. {\bf B~522}, 321 (1998).

\bibitem{Griesshammer}
H.W.~Griesshammer, Phys. Rev. {\bf D~58}, 094027 (1998);
preprints hep-ph/9804251, hep-ph/9810235.

\bibitem{GSS}
J.~Gasser, M.E.~Sainio, and A.~\v{S}varc, Nucl. Phys. {\bf B~307}, 779
(1988).

\bibitem{HBChPT}
E.~Jenkins and A.V.~Manohar, Phys. Lett {\bf B~255}, 558 (1991);
V.~Bernard, N.~Kaiser, J.~Kambor, and U.-G.~Meissner,
Nucl. Phys. {\bf B~388}, 315 (1992);
P.J.~Ellis and H.-B.~Tang, Phys. Rev. {\bf C~57}, 3356 (1998).

\bibitem{Becher}
T.~Becher and H.~Leutwyler, Eur. Phys. J. {\bf C~9}, 643 (1999).

\bibitem{NN}
S.~Weinberg, Phys. Lett. {\bf B~251}, 288 (1990); Nucl. Phys. {\bf B~363}, 3
(1991); Phys. Lett. {\bf B~295}, 114 (1992);
D.B.~Kaplan, M.J.~Savage and M.B.~Wise, Nucl. Phys. {\bf B~534}, 329 (1998);
Phys. Lett. {\bf B~424}, 390 (1998).

\bibitem{Feshbach}
H.~Feshbach, Ann. Phys. {\bf 5}, 357 (1958); {\bf 19}, 287 (1962).

\bibitem{Schwinger}
J.~Schwinger, J. Math. Phys. {\bf 5}, 1606 (1964).

\bibitem{Bogoliubov}
N.N.~Bogoliubov and D.V.~Shirkov, ``Introduction to the Theory of Quantized
Fields'', New-York - London - Sydney, John Wiley \& Sons, (1959).

\bibitem{Bunatian}
G.G.~Bunatian, Nucl. Phys. {\bf A~645}, 314 (1999).

\bibitem{Gall}
A.~Gall, PhD thesis, University of Bern (1998), preprint BUTP-99-19,
hep-ph/9910364.

\bibitem{EOM1}
K.~Hasebe and Y.~Sumino, preprint hep-ph/9910424.

\bibitem{EOM2}
D.B.~Kaplan, M.J.~Savage, and M.B.~Wise, Phys. Rev. {\bf C~59}, 617 (1999).


\end{thebibliography}
\end{document}